\documentclass[journal]{IEEEtran}
\usepackage{tikz}
\usetikzlibrary{decorations.pathmorphing, shapes}
\usetikzlibrary{positioning}
\usetikzlibrary{calc}
\usetikzlibrary{arrows}
\usepackage{relsize}
\usepackage[americanvoltage]{circuitikz}

\ifCLASSINFOpdf
\else
\fi

\usepackage{amsmath}
\usepackage{amssymb}
\usepackage{pgfplots}
\usetikzlibrary{calc}
\usepackage[americanvoltage]{circuitikz}
\usepackage{paralist}

\usepackage[caption=false]{subfig}

\newcommand{\LArgtwo}[2]{\left[\vec{\cal L}_{#2} #1 \right](\vec{r})}
\newcommand{\KArgtwo}[2]{\left[\vec{\cal K}_{#2} #1 \right](\vec{r})}
\newcommand{\vr}{\vec{r}}
\newcommand{\vrp}{\vec{r}\,'}

\newcommand{\red}[1]{{\color{red} #1}}            
\newcommand{\blue}[1]{{\color{blue} #1}}            
\newcommand{\green}[1]{{\color{magenta} #1}}            
            
\newcommand{\brown}[1]{{\color{brown} #1}}

\newcommand{\vect}[1]{\mathbf{#1}}

\newcommand{\matr}[1]{\mathbf{#1}}

\newcommand{\junk}[1] {}

\def\XXint#1#2#3{{\setbox0=\hbox{$#1{#2#3}{\int}$}
\vcenter{\hbox{$#2#3$}}\kern-.5\wd0}}

\newcommand*\widebar[1]{%
  \hbox{%
    \vbox{%
      \hrule height 0.5pt 
      \kern0.3ex
      \hbox{%
        \kern-0.05em
        \ensuremath{#1}%
        \kern-0.05em
      }%
    }%
  }%
}

\usetikzlibrary{patterns}

\title{A Fast Macromodeling Approach to Efficiently Simulate Inhomogeneous Electromagnetic Surfaces}
\author{Utkarsh~R.~Patel,~\IEEEmembership{Member,~IEEE}, 
        Piero~Triverio,~\IEEEmembership{Senior Member,~IEEE}, and
Sean V.~Hum,~\IEEEmembership{Senior Member,~IEEE}\\[12pt]
Paper submitted to the IEEE Transactions on Antennas and Propagation
\thanks{Manuscript received ...; revised ...}%
\thanks{U.~R.~Patel, P.~Triverio, and S. V.~Hum are with the Edward S. Rogers Sr. Department of Electrical and Computer Engineering, University of Toronto, Toronto, M5S 3G4 Canada (email: utkarsh.patel@mail.utoronto.ca, piero.triverio@utoronto.ca, sean.hum@utoronto.ca).}
\thanks{This project was supported by NSERC Strategic Partnership Grant for Projects}
}

\begin{document}

\maketitle
\begin{abstract}
The full-wave simulation of complex electromagnetic surfaces such as reflectarrays and metasurfaces is a challenging problem.
In this paper, we present a macromodeling approach to efficiently simulate complex electromagnetic surfaces composed of PEC traces, possibly with fine features, on a finite-sized multilayer dielectric substrate.
In our approach, we enclose each element of the structure with a fictitious surface.
By applying the equivalence principle on each surface, we derive a macromodel for each element of the array.
This macromodel consists of a linear operator that relates the equivalent electric and magnetic current densities introduced on the fictitious surface. 
Mutual coupling between the elements of the structure is captured by the equivalent current densities in a fully accurate way.
The crux of the proposed technique is to solve for equivalent current densities on the fictitious surface instead of directly solving for the actual current densities on the original scatterer. When simulating complex surfaces, this approach leads to fewer unknowns and better conditioning.
We also propose a rigorous acceleration algorithm based on the fast Fourier transform to simulate electrically large surfaces.
Numerical results demonstrate that the proposed approach is significantly faster and requires less memory than commercial solvers based on the surface integral equation method, while giving accurate results. 
\end{abstract}

\section{Introduction}

Electromagnetic (EM) simulation tools are necessary to design complex EM surfaces, such as metasurfaces, reflectarrays, and transmitarrays, that are used in many communication and imaging applications.
Most complex EM surfaces have periodic spacing between their elements. However, these surfaces are typically inhomogeneous with distinct unit cells, each of which is  designed to locally manipulate amplitude and phase of an incident EM wave.
Most EM surfaces are electrically large with dimensions of tens to hundreds of wavelengths and are composed of stacked layers of conductor traces on an electrically-thin dielectric substrate.
Furthermore, some of these surfaces possess electrically fine features.
Due to the complexity of these surfaces, analysing them with a full-wave simulation requires solving a large number of unknowns, involving prohibitive amounts of memory and CPU time.

Due to the aforementioned challenges, most designers simulate each unit cell of an EM surface with periodic boundary conditions~\cite{Bhattacharyya2006}\nocite{wan1995}--\cite{jin2014finite}.
Radiation from the array is then computed with array factor analysis~\cite{Zhou2011}\nocite{ Arrebola2008}--\cite{Li2009_Array}.
This technique is demonstrably faster than performing a full-wave simulation of the entire array. 
However, it does not accurately model the mutual coupling between dissimilar unit cells. 
Furthermore, that particular approach does not accurately capture edge effects due to the finite size of the EM surface. 
Hence, this technique cannot accurately predict directivity, side lobe levels, and null locations of a typical EM surface with abrupt changes between adjacent unit cells. 

Among full-wave techniques, the surface integral equation (SIE) method~\cite{Gibson2009}--\cite{ Chew2008} is commonly used to simulate scattering problems.
EM surfaces can be simulated with the SIE method using either the multilayer Green's function~(MLGF)~\cite{Mic97} or the equivalence principle-based PMCHWT formulation~\cite{Poggio1973}\nocite{Chang1977}--\cite{Wu1977}.
In MLGF-based formulations, the surface current density on all conductor traces inside a unit cell is discretized and solved using the method of moments. 
The dielectric substrate supporting the surface is assumed to be infinitely wide and is modeled with the multilayer Green's function~\cite{Bailey1982}\nocite{Pozar1987}--\cite{Mosig1988}.
Computation time to solve the linear system can be significantly reduced by using iterative techniques such as conjugate gradient descent and the generalized minimal residual~(GMRES) methods instead of a direct method based on the LU factorization.
Electrically large composite objects can be simulated with the fast multipole method (FMM)~\cite{Coifman1993}--\cite{Greengard1988}, the multilevel fast multipole method (MLFMM)~\cite{MLFMM} or the adaptive integral method (AIM)~\cite{Bleszynski96}--\nocite{Yuan2003,Wang1998,Bindiganavale98,Zhuang1996}\cite{Zhu2005}.
The convergence of iterative solvers may also be improved via efficient preconditioners~\cite{Andriulli2008}.
While acceleration methods reduce memory consumption and solution time, they do not reduce the number of unknowns.
Reduced-order techniques such as those employing macrobasis functions~\cite{Sut00},  characteristic basis functions~\cite{Prakash2003}, synthetic basis functions~\cite{Mat07}, and eigencurrent basis functions~\cite{Bekers2009} can reduce the number of unknowns by projecting field quantities onto a new set of basis functions that are obtained via an eigenvalue or singular value decomposition. 
While this approach is efficient for simple unit cell geometries, the cost of an eigenvalue or singular value decomposition for a complex unit cell can be quite high.

The MLGF-based formulation does not accurately model edge effects and also ignores spillover loss, which results in discrepancies in the radiation pattern. For accurate results, the PMCHWT formulation may be applied to model dielectric substrates with equivalent electric and magnetic current densities~\cite{Chew2008}. 
In this technique, the homogeneous 3-D Green's functions of free space and dielectric media are used to compute fields radiated by equivalent electric and magnetic current densities~\cite{Har61}. 
This idea was later generalized to simulate composite objects made up of dielectrics and PECs~\cite{Oijala2005}--\cite{Oijala2005_2}.
While the PMCHWT formulation can improve the accuracy of the simulated results, solving for equivalent electric and magnetic currents on the boundary of each dielectric layer  significantly increases the unknowns count.
Moreover, multiscale features that are present in complex unit cells may also slow down convergence.

Domain decomposition methods~(DDMs) comprise yet another class of techniques that can efficiently simulate electrically-large problems~\cite{DDM_Lee2005}--\cite{Peng2011}.
The equivalence principle algorithm (EPA) is a type of DDM approach suitable to tackle multiscale electromagnetic problems~\cite{li2007}\nocite{Lancellotti2009, Oijala2009}--\cite{Patel2018}.
In the EPA, a complex electromagnetic scatterer is enclosed by a fictitious surface. 
The Love's equivalence principle and the SIE method are then applied to derive a scattering operator that relates incident and scattered EM fields on the fictitious surface. 
The inter-element coupling is captured by the so-called translation operator.
Since the EPA requires solving for unknowns only on the equivalent surfaces, it requires fewer unknowns and typically has better convergence properties.
To tackle large problems, the EPA has been combined with  higher-order basis functions~\cite{Li2008}--\cite{Shao2013} and with acceleration algorithms such as the MLFMM~\cite{Shao2011}.

This paper proposes a macromodeling technique to simulate complex electromagnetic surfaces, such as reflectarrays and metasurfaces, composed of PEC traces on a multilayer dielectric substrate. 
The contribution in this paper is threefold. 
First, the macromodeling approach is presented in this paper to efficiently model complex unit cells. 
Despite the fact that the proposed technique is based on the equivalence principle like the EPA, the two techniques are formulated differently. 
In the past, the EPA was applied to simulate antenna arrays with non-zero spacing between array elements~\cite{Li2008}, \cite{Oijala2009}--\cite{Patel2018},~\cite{Shao2011}--\cite{Xiang2013}.
In all these works, the EPA was developed for scatterers that can be fully enclosed by a fictitious surface. 
However, when simulating electromagnetic surfaces, the fictitious surfaces must traverse the layered substrate. 
Furthermore, fictitious surface enclosing adjacent elements partially overlap. 
In the proposed macromodeling approach, we discretize the electric and magnetic field integral equations for each region inside a fictitious surface, impose electromagnetic boundary conditions using the PMCHWT formulation, and then eliminate unknowns associated with field quantities inside the fictitious surface using the Schur complement. 
The macromodeling approach is developed to simulate electromagnetic surfaces, hence, in contrast to previous approaches, the proposed method can deal with:
\begin{itemize}
\item fictitious surfaces that traverse multilayer dielectric substrate;
\item overlapping fictitious surfaces of adjacent unit cells;
\item fictitious surfaces that are backed by a ground plane, as in the case of reflectarrays.  
\end{itemize}
Junctions that are formed at intersections between two fictitious surfaces or between a fictitious surface and a layered substrate are properly modeled with the PMCHWT formulation~\cite{Oijala2005_2}.
The second contribution of this paper is a rigorous FFT-based acceleration method to simulate electrically-large arrays. After instantiating a macromodel for each element of the array, we can effectively model an array of inhomogeneous unit cells with an array of indentically-meshed equivalent current densities. 
This allows us to exploit the Toeplitz structure of the discretized integral equation matrices to compute inter-element coupling accurately and efficiently via FFT. 
The proposed Toeplitz approach is more efficient than AIM because it does not require computing precorrections or decomposing far-field terms into scalar and vector potentials. 
Previously, this FFT-based acceleration approach was proposed to simulate an array of identical antenna elements that are separated by a finite distance in an FEM-IE hybrid solver~\cite{Kin03}. However, to the best of our knowledge, it has not been applied to simulate an array of dissimilar elements.
Finally, to the best of our knowledge, an EPA technique has never been applied to simulate electromagnetic surfaces realized on a layered substrate. In this paper, we use the proposed technique to simulate two practical electromagnetic structures and compare the results against other commercial solvers and measurement results.

This paper is organized as follows. 
In Sec.~\ref{sec:ch5_macromodel_composite_macromodel_single}, we discuss how to generate a macromodel for a single element of an array. 
Then, in Sec.~\ref{sec:ch5_macromodel_composite_equivalent_exterior}, we discuss how to simulate an entire EM surface using an array of macromodels. 
To simulate large arrays, we present an FFT-based acceleration algorithm in Sec.~\ref{sec:ch5_macromodel_composite_toeplitz}. 
Finally, in Sec.~\ref{sec:Results}, we present examples to validate the proposed technique against other numerical methods and experimental measurements.

\begin{figure}[t]
\centering
\includegraphics[width=0.8\columnwidth, bb=0 0 900 500]{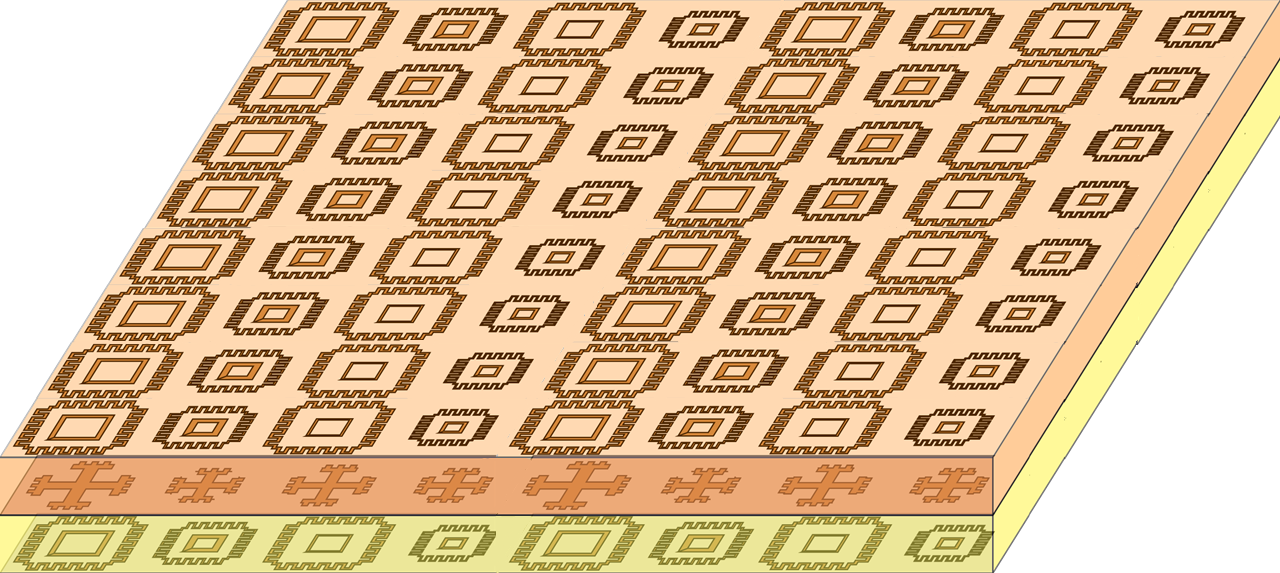}
\caption{Sample EM surface composed of a two-layer dielectric substrate (shown in yellow and orange) and metallic traces (shown in brown).}
\label{fig:sample_surface}
\end{figure}

\section{Macromodel Generation}
\label{sec:ch5_macromodel_composite_macromodel_single}

We consider the problem of computing scattering from a complex EM surface, such as the one shown in Fig.~\ref{fig:sample_surface}.
We assume that the surface is inhomogeneous, hence the PEC traces in each unit cell may have different sizes and shapes to locally control reflection and transmission coefficients.
We assume that PEC traces on two adjacent unit cells are not connected. Each unit cell may have a PEC ground plane at the bottom of the dielectric substrate.

\subsection{Discretization}
\label{sec:ch5_macromodel_composite_problem_def}

\begin{figure}[t]
\centering
\null \hfill
\subfloat[\label{fig:sample_cell_equivalent_1} Original setup]{\includegraphics{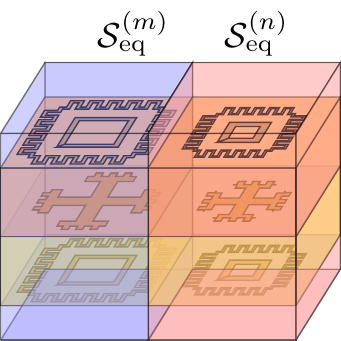}}
\hfill
\subfloat[\label{fig:sample_cell_equivalent_2} Equivalent setup]{\includegraphics{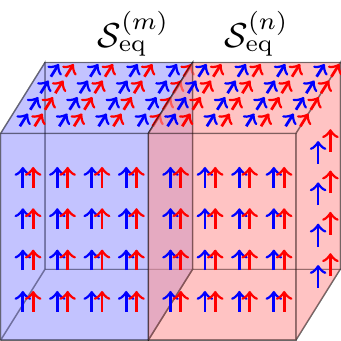}}
\hfill \null
\caption{(a): Original configuration: two unit cells of the array in Fig.~\ref{fig:sample_surface} are enclosed by fictitious closed surfaces (shown in red and blue). (b): Equivalent configuration: unit cells are modeled by equivalent electric and magnetic current densities that are introduced on closed surfaces.}
\end{figure}

In order to generate a macromodel, we enclose the $m$-th unit cell with a closed surface ${\cal S}_{\mathrm{eq}}^{(m)}$ such that all PEC traces of the $m$-th element are inside ${\cal S}_{\mathrm{eq}}^{(m)}$. Furthermore, the entire layered substrate is enclosed by the union of all fictitious surfaces.
Some portions of the sidewalls of ${\cal S}_{\mathrm{eq}}^{(m)}$ may traverse the dielectric substrate.
Since we are simulating EM surfaces with uniform spacing between all elements, it is convenient to use the surface of a rectangular prism with dimensions of the unit cell as the fictitious surface.
If the electromagnetic structure is backed by a PEC ground plane, then the bottom face of ${\cal S}_{\mathrm{eq}}^{(m)}$ is set to coincide with the ground plane. 
Fig.~\ref{fig:sample_cell_equivalent_1} shows this step for two of the elements in Fig.~\ref{fig:sample_surface}. 
Next, we mesh all surfaces on and inside ${\cal S}_{\mathrm{eq}}^{(m)}$ with triangular elements. 
Since the fictitious surface enclosing adjacent unit cells may partially overlap, we mesh opposite side faces of ${\cal S}_{\mathrm{eq}}^{(m)}$ with identical meshes in order to be able to, later on, properly enforce EM boundary conditions.

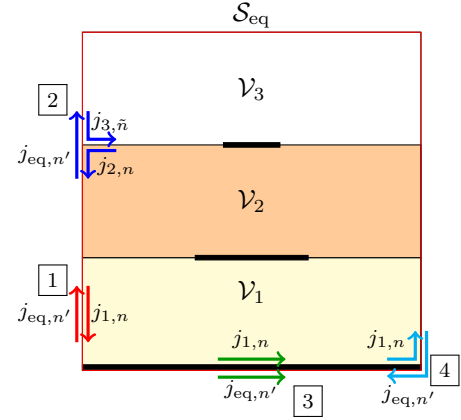
\begin{figure}[t]
\centering
\begin{tikzpicture}[scale = 1.5]

  \draw[fill = yellow!20] (0,0) rectangle (3, 1);
  \draw[fill = orange!40] (0,1) rectangle (3, 2);
  \draw[fill = black] (1,0.98) rectangle (2, 1.02);
  \draw[fill = black] (1.25,1.98) rectangle (1.75, 2.02);
\draw[fill=black] (0,0.01) rectangle (3,0.05);
  \draw (0,0) rectangle (3,3);
  
  \draw[red] (0,0) -- (0,3);
  \draw[red] (3,0) -- (3,3);
  \draw[red] (0,0) -- (3,0);
  \draw[red] (0,3) -- (3,3);
    \node at (1.5,0.7) {${\cal V}_1$};
  \node at (1.5,1.5) {${\cal V}_2$};
  \node at (1.5,2.5) {${\cal V}_3$};



	\node at (1.5, 0.1) [above] {\footnotesize $j_{1,n}$};

\node at (0.00, 0.5) [right] {\footnotesize $j_{1,n}$};
\node at (0.00, 0.5) [left] {\footnotesize $j_{\mathrm{eq},n'}$};
\node at (2.7, 0.1) [above] {\footnotesize $j_{1,n}$};
\node at (2.9, -0.05) [below] {\footnotesize $j_{\mathrm{eq},n'}$};
\draw [blue, very thick, ->] (0.05, 2.3) -- (0.05, 2.05) -- (0.3,2.05);
\draw [blue,very thick, <- ] (0.05, 1.7) -- (0.05, 1.95) -- (0.3,1.95);
\draw [blue, very thick,<-] (-0.05, 2.3) -- (-0.05,1.7);
\node at (-0.05,2.4) [left] {\footnotesize \boxed{2}};
\node at (0,1.9) [left] {\footnotesize $j_{\mathrm{eq},n'}$};
\node at (0.05,1.8) [right] {\footnotesize $j_{2,{n}}$};
\node at (0,2.2) [right] {\footnotesize $j_{3,\tilde{n}}$};

\draw [red, very thick,->] (-0.05, 0.25) -- (-0.05,0.75);
\draw [red, very thick, <-] (0.05, 0.25) -- (0.05,0.75);

\node at (-0.05,0.8) [left] {\footnotesize \boxed{1}};

\draw [color=black!40!green, very thick, ->] (1.2, 0.1) -- (1.8,0.1);
\draw [color=black!40!green, very thick, ->] (1.2, -0.06) -- (1.8,-0.06);

\node at (2,-0.05) [below] {\footnotesize \boxed{3}};
\node at (1.5,0) [below] {\footnotesize $j_{\mathrm{eq},n'}$};

\draw [cyan, very thick,-> ] (2.7,0.1) -- (2.95,0.1) -- (2.95,0.35);
\draw [cyan, very thick, <- ] (2.7,-0.05) -- (3.05,-0.05) -- (3.05,0.35);

\node at (3.0,0) [right] {\footnotesize \boxed{4}};
\node  at (1.5,3.15) {${\cal S}_{\mathrm{eq}}$};
\end{tikzpicture}
\caption{Side view of a sample unit cell. Fictitious surface ${\cal S}_{\mathrm{eq}}$ enclosing the unit cell is drawn in red. Rest of the surfaces (in black) are interior surfaces. Regions ${\cal V}_1$ and ${\cal V}_2$ are dielectric regions.  ${\cal V}_3$ is an air region that is introduce in order to ensure that all PEC surfaces (except for the ground plane) are strictly inside ${\cal S}_{\mathrm{eq}}$. Special junctions are shown in red, blue, green, and cyan colors and labeled {\footnotesize \boxed{1}}, $\hdots$, {\footnotesize \boxed{4}}.}
\label{fig:composite}
\end{figure}

Throughout the rest of this section, we focus on creating a macromodel for the $m$-th element of the array.  For brevity, we omit the superscript $(m)$ from all geometrical and field quantities for now. These superscripts will be reintroduced when we capture mutual coupling between multiple elements. 
To explain macromodel generation, let us consider the cross-section of a sample unit cell with a PEC ground plane shown in Fig.~\ref{fig:composite}. 
This unit cell has $V=3$ homogeneous regions. The top region in this case is air, so that all PEC surfaces (excluding the PEC ground plane) are strictly inside ${\cal S}_{\mathrm{eq}}$. 
The $v$-th region is denoted by ${\cal V}_v$.
The surface enclosing the $v$-th region is denoted by ${\cal S}_{v}$. 

According to the equivalence principle~\cite{Chew2008}, electromagnetic fields inside or outside the $v$-th region can be computed through equivalent electric and magnetic current densities
\begin{subequations}
\begin{align}
\vec{J}_v(\vr) &= \hat{n}_v \times \vec{H}_v(\vr) \label{eq:Heq}\\
\vec{M}_v(\vr) &= -\hat{n}_v\times \vec{E}_v(\vr) \label{eq:Eeq}
\end{align}
\end{subequations}
that are introduced on ${\cal S}_{v}$. In~\eqref{eq:Heq}--\eqref{eq:Eeq}, $\vec{H}_v(\vr)$ and $\vec{E}_v(\vr)$ are the magnetic and electric fields tangential to ${\cal S}_{v}$, and $\hat{n}_{v}$ is the unit normal vector pointing into the $v$-th region. Throughout the rest of this paper, we will omit the subscript in $\hat{n}_{v}$ for brevity.
We expand the equivalent electric and magnetic current densities on all surfaces inside ${\cal S}_{\mathrm{eq}}$ with RWG basis functions.
For $\vec{r} \in {\cal S}_{v}$, the electric and magnetic current densities are expanded as
\begin{subequations}
\begin{align}
\vec{J}_{v}(\vr) &= \sum_{n=1}^{N_{v}} j_{v,n} \vec{\Lambda}_{v,n}(\vr) \label{eq:Hdiscrete}\\
\vec{M}_{v}(\vr) &= \sum_{n=1}^{N_{v}'} m_{v,n} \vec{\Lambda}_{v,n}(\vr) \label{eq:Ediscrete}\,,
\end{align}
\end{subequations}
where $\vec{\Lambda}_{v,n}$ is the $n$-th RWG basis function~\cite{RWG} on ${\cal S}_v$, and $N_v$ and $N_v'$ are, respectively, the total number of RWG functions used to expand the electric and magnetic current densities on ${\cal S}_v$. 
 
If any subsurface of ${\cal S}_{v}$ is a PEC, then the equivalent electric current density on it is the same as the surface electric current density. On the other hand, the equivalent magnetic current density on a PEC is zero, and is thus not discretized. Hence, $N_{v}$ is always greater than or equal to $N_{v}'$. All electric current density coefficients on the surface enclosing the $v$-th region are collected into a vector $\vect{J}_{v} = \begin{bmatrix} j_{v,1} & \hdots & j_{v,N_v}\end{bmatrix}^T$ and all magnetic current density coefficients are collected into a vector $\vect{M}_{v} = \begin{bmatrix} m_{v,1} & \hdots & m_{v,N_v'}\end{bmatrix}^T$. 

We also expand the equivalent electric and magnetic current density for $ \vec{r} \in {\cal S}_{\mathrm{eq}}$ with RWG basis functions
\begin{subequations}
\begin{align}
 \vec{J}_{\mathrm{eq}}(\vr) &= \sum_{n=1}^{N_{\mathrm{eq}}} j_{\mathrm{eq},n} \vec{\Lambda}_{\mathrm{eq},n}(\vr) \label{eq:Hdiscretebox}\\
  \vec{M}_{\mathrm{eq}}(\vr) &= \sum_{n=1}^{N_{\mathrm{eq}}'} m_{\mathrm{eq},n} \vec{\Lambda}_{\mathrm{eq},n}(\vr) \label{eq:Ediscretebox}\,,
\end{align}
\end{subequations}
where the unit normal vector $\hat{n}$ points in the outer region (free space), $N_{\mathrm{eq}}$ and $N_{\mathrm{eq}}'$ are the number of RWG basis functions used to discretize the electric and magnetic current densities, respectively. 
The electric and magnetic current density coefficients in~\eqref{eq:Hdiscretebox}--\eqref{eq:Ediscretebox} are collected into vectors $\vect{J}_{\mathrm{eq}} = \begin{bmatrix} j_{\mathrm{eq},1} & \hdots & j_{\mathrm{eq},N_{\mathrm{eq}}} \end{bmatrix}^T$ and $\vect{M}_{\mathrm{eq}} = \begin{bmatrix}  m_{\mathrm{eq},1} & \hdots & m_{\mathrm{eq},N_{\mathrm{eq}}'}\end{bmatrix}^T$, respectively. Furthermore, we collect electric and magnetic current density coefficients into a vector 
$\widetilde{\vect{X}}_{\mathrm{eq}} = \begin{bmatrix} \vect{J}_{\mathrm{eq}}^T & \vect{M}_{\mathrm{eq}}^T \end{bmatrix}^T$.

\subsection{Surface Integral Equations}

We now apply the SIE method to generate a  macromodel for the $m$-th element of the array. 
The macromodel will ultimately capture electromagnetic scattering from a complex constituent scatterer, such as the one shown in Fig.~\ref{fig:sample_cell_equivalent_1}, with equivalent current densities on ${\cal S}_{\mathrm{eq}}$,  as shown in Fig.~\ref{fig:sample_cell_equivalent_2}.

According to Love's equivalence principle~\cite{Chew2008}, we can relate the equivalent electric and magnetic current densities on ${\cal S}_{v}$ through the electric field integral equation (EFIE) and the magnetic field integral equation (MFIE)
\begin{subequations}
\begin{align}
-\hat{n} \times \vec{M}_{v}(\vr) &=  -j\omega\mu_0 \hat{n} \times \hat{n} \times  \LArgtwo{\vec{J}_{v}(\vrp)}{v} \nonumber \\
& - \hat{n} \times \hat{n} \times \KArgtwo{\vec{M}_{v}(\vrp)}{v}\,, \label{eq:SIE_TEFIE1_Composite}\\
\hat{n} \times \vec{J}_{v}(\vr) &= -j\omega\varepsilon_0 \hat{n} \times \hat{n} \times \LArgtwo{\vec{M}_{v}(\vrp)}{v}  \nonumber \\
& + \hat{n} \times \hat{n} \times \KArgtwo{\vec{J}_{v}(\vrp)}{v}\,, \label{eq:SIE_TMFIE1_Composite}
\end{align}
\end{subequations}
where operators $\hat{n} \times \vec{\cal L}_v$ and $\hat{n} \times \vec{\cal K}_v$ are given by
\begin{align}
\hat{n} \times \LArgtwo{\vec{X}(\vrp)}{v} &= \hat{n} \times \left[1 + \frac{\nabla \nabla \cdot}{k_v^2}  \right] \int_{{V}} G_v(\vr,\vrp) \vec{X}(\vrp) dV\,' \label{eq:SIE_LOperator1} \\
\hat{n} \times \KArgtwo{\vec{X}(\vrp)}{v} &= \hat{n} \times \mathrm{p.v.} \left[ \nabla \times \int_{V} G_v(\vr,\vrp) \vec{X}(\vrp) dV\,'\right] \nonumber \\
&\quad + \frac{\vec{X}(\vr)}{2} \,. \label{eq:SIE_KOperator1}
\end{align}
Operators $\vec{\cal L}_v$ and $\vec{\cal K}_v$ in~\eqref{eq:SIE_LOperator1}--\eqref{eq:SIE_KOperator1} are evaluated with the wave number $k_v = \omega \sqrt{\mu_0 \varepsilon_v}$, electrical permittivity $\varepsilon_v$, and homogeneous Green's function $G_v(\vr,\vrp)$ of the $v$-th region. Operator $\mathrm{p.v.}$ stands for the principal value.

Next, we substitute~\eqref{eq:Hdiscrete}--\eqref{eq:Ediscrete} into~\eqref{eq:SIE_TEFIE1_Composite}--\eqref{eq:SIE_TMFIE1_Composite} and test the resulting integral equations with RWG basis functions.
For the $v$-th region, we obtain the following linear system of equations
\begin{equation}
\begin{bmatrix}
\matr{L}_{v}^{E} & \matr{K}_{v}^{E} \\ \matr{K}_{v}^H & \matr{L}_{v}^H \end{bmatrix} \begin{bmatrix} \vect{J}_{v} \\ \vect{M}_{v}  \end{bmatrix} = \begin{bmatrix} \vect{0} \\ \vect{0} \end{bmatrix} \,,
\label{eq:ch5_macromodel_composite_SIE_Block}
\end{equation}
where matrices $\matr{L}_v^E$ and $\matr{L}_v^H$ are obtained by discretizing the $\vec{\cal L}_v$ operator in~\eqref{eq:SIE_TEFIE1_Composite} and~\eqref{eq:SIE_TMFIE1_Composite}, respectively. Likewise, $\matr{K}_v^E$ and $\matr{K}_v^H$ are obtained by discretizing the $\vec{\cal K}_v$ operator in~\eqref{eq:SIE_TEFIE1_Composite} and~\eqref{eq:SIE_TMFIE1_Composite}, respectively.

We now collect~\eqref{eq:ch5_macromodel_composite_SIE_Block} for all $V$ regions into a larger system of linear equations
\begin{align}
\underbrace{\left[ \begin{array}{c|c|c} \begin{matrix} \matr{L}_{1}^{E} & \matr{K}_{1}^{E} \\ \matr{K}_{1}^H & \matr{L}_{1}^H \end{matrix} & 0 & 0 \\ \hline
0 & \ddots & 0 \\ \hline
0 & 0 & \begin{matrix} \matr{L}_{V}^{E} & \matr{K}_{V}^{E} \\ \matr{K}_{V}^H & \matr{L}_{V}^H \end{matrix} \end{array} \right]
}_{\matr{Z}}
\underbrace{\left[ \begin{array}{c}
 \vect{J}_{1} \\ \vect{M}_{1}  \\ \hline \vdots \\ \hline \vect{J}_{V} \\ \vect{M}_{V}  \end{array}\right] }_{\vect{X}} = 
\left[ \begin{array}{c}  \vect{0} \\ \vect{0}  \\ \hline \vdots \\ \hline \vect{0} \\ \vect{0} 
\end{array} \right] \,.
\label{eq:discretized_SIE}
\end{align}
To simplify the presentation of subsequent sections, we will denote the block-diagonal matrix in~\eqref{eq:discretized_SIE} with $\matr{Z}$ and the vector of field coefficients with $\vect{X}$.
Note that the right-hand side of~\eqref{eq:discretized_SIE} is zero due to the absence of any sources inside ${\cal S}_{\mathrm{eq}}$.

\subsection{Enforcement of Boundary Conditions}
\label{sec:ch5_macromodel_composite_BC_Unitcell}

Since we discretized equivalent electric and magnetic current densities on either side of an interface or a junction, we have redundant unknowns.
We enforce boundary conditions, such as the continuity of electromagnetic fields across the interface of two homogeneous regions,
on junctions and interfaces inside ${\cal S}_{\mathrm{eq}}$ to eliminate redundant unknowns.
Suppose that following the removal of redundant unknowns, the final set of unknowns for the $m$-th element is collected into a vector 
\begin{equation} 
\widetilde{\vect{X}} = \begin{bmatrix} \widetilde{\vect{X}}_{\mathrm{eq}}^T & \widetilde{{\vect{X}}}_{\mathrm{int}}^T \end{bmatrix}^T \,,
\end{equation}
where $\widetilde{\vect{X}}_{\mathrm{eq}}$ collects unknown current coefficients on ${\cal S}_{\mathrm{eq}}$ that appear in~\eqref{eq:Hdiscretebox}--\eqref{eq:Ediscretebox} and $\widetilde{\vect{X}}_{\mathrm{int}}$ collects the rest of the unknowns associated with current densities inside ${\cal S}_{\mathrm{eq}}$. 
We relate $\vect{X}$ to $\widetilde{\vect{X}} $ by
\begin{align}
\matr{X} = \matr{U} \widetilde{\matr{X}}\,,
\label{eq:incidence1}
\end{align}
where $\matr{U}$ is a sparse matrix with a few entries per row, whose elements in turn may be $\pm 1$. This matrix serves two purposes. First,  it eliminates redundant unknowns by explicitly enforcing continuity of tangential electric and magnetic fields on the interfaces between two or more regions.  
Second, it rearranges the list of unknowns in order to group unknowns on ${\cal S}_{\mathrm{eq}}$ and unknowns inside ${\cal S}_{\mathrm{eq}}$.

To discuss how to enforce all boundary conditions, we reconsider the sample unit cell shown in Fig.~\ref{fig:composite}.
All boundary conditions can be classified into two sets: 
boundary conditions for interfaces and junctions inside ${\cal S}_{\mathrm{eq}}$, whose enforcement has been well-discussed in the literature~\cite{Oijala2005_2}; and boundary conditions for interfaces and junctions on ${\cal S}_{\mathrm{eq}}$, whose enforcement is discussed in this paper.
Let us first summarize which boundary conditions have to be enforced on fields inside ${\cal S}_{\mathrm{eq}}$. 
These boundary conditions are discussed in detail in other works~\cite{Oijala2005_2, Gibson2009}, however, our goal is to discuss their enforcement using $\matr{U}$:
\begin{enumerate}
\item {\bf Interface of dielectric regions:} The tangential electric and magnetic fields are continuous across the interface between two dielectric regions $v$ and $v'$. Hence, expansion coefficients  in~\eqref{eq:Hdiscrete}--\eqref{eq:Ediscrete} are set to satisfy $j_{v,n} = j_{v',n'}$ and $m_{v,n} = m_{v',n'}$, assuming the RWG basis functions $\vec{\Lambda}_{v,n}$ and $\vec{\Lambda}_{v',n'}$ are co-located but oriented in opposite directions.
We can enforce the continuity of the tangential magnetic field by collecting, for example, $j_{v,n}$ into $
\widetilde{\vect{X}}_{\mathrm{int}}$. 
Then, entries $(q^{j_{v,n}}, \tilde{q}^{j_{v,n}})$ and $(q^{j_{v',n'}},\tilde{q}^{j_{v,n}})$ of $\matr{U}$ are set to $1$, where $q^{\alpha}$ and $\tilde{q}^{\beta}$ are, respectively, the index of the entries associated to coefficient $\alpha$ in $\vect{X}$ and $\beta$ in $\widetilde{\vect{X}}$.
The continuity of tangential electric field can be enforced similarly.

\item {\bf PEC surface at the interface of two regions:} The tangential magnetic fields on the two sides of a PEC interface between regions $v$ and $v'$ are independent. Hence, we keep two unknowns ($j_{v,n}$ and $j_{v',n'}$) into $\widetilde{\vect{X}}_{\mathrm{int}}$. In this case, entries $(q^{j_{v,n}},\tilde{q}^{j_{v,{n}}})$ and $(q^{j_{v',{n'}}},\tilde{q}^{j_{v',{n'}}})$ of $\matr{U}$ are set to $1$. The tangential electric field is zero on the PEC surface and, hence, was not discretized in~\eqref{eq:Ediscrete}.
\item {\bf PEC-dielectric junctions:} A PEC-dielectric junction is defined on the boundary of a PEC surface, where half of the RWG basis function is on the PEC surface and the other half is on the interface between dielectrics. Due to continuity of electromagnetic fields, the tangential magnetic field is continuous on the interface between regions $v$ and $v'$. Therefore, the electric current density coefficients are related as $j_{v,n} = j_{v',n'}$. We collect $j_{v,n}$ into the vector $\widetilde{\vect{X}}_{\mathrm{int}}$. To enforce this boundary condition, entries $(q^{j_{v,n}}, \tilde{q}^{j_{v,n}})$ and $(q^{j_{v',n'}}, \tilde{q}^{j_{v,n}})$ of $\matr{U}$ are set  to $1$.
\end{enumerate}
As discussed above, after enforcing the above boundary conditions, all unique unknowns inside ${\cal S}_{\mathrm{eq}}$ will appear in $\widetilde{\vect{X}}_{\mathrm{int}}$. The following boundary conditions need to be enforced on ${\cal S}_{\mathrm{eq}}$:
\begin{enumerate}
\item {\bf Interface on ${\cal S}_{\mathrm{eq}}$ between the outer region and an inner region:} We consider a sample interface on ${\cal S}_{\mathrm{eq}}$ between the outer region and inner region ${\cal V}_1$ that is shown in the region labeled with {\footnotesize \boxed{1}} in Fig.~\ref{fig:composite}. On this interface, due to continuity of the tangential magnetic field, $j_{1,n}$ is equal to $j_{\mathrm{eq},n'}$ for some values of $n$ and $n'$, assuming that basis functions $\vec{\Lambda}_{1,n}$ and $\vec{\Lambda}_{\mathrm{eq},n'}$ are oriented in the direction shown by the red arrows in Fig.~\ref{fig:composite}. 
This boundary condition can be enforced by setting entries $(q^{j_{1,n}}, \tilde{q}^{{j}_{\mathrm{eq},n'}})$ and $(q^{j_{\mathrm{eq},n'}}, \tilde{q}^{{j}_{\mathrm{eq},n'}})$ of $\matr{U}$ to $1$.
Continuity of the tangential electric field can be enforced similarly.
\item {\bf Junction on ${\cal S}_{\mathrm{eq}}$ between two interior regions and the outer region:} We consider the sample junction on ${\cal S}_{\mathrm{eq}}$ that is shown in the region labeled with {\footnotesize \boxed{2}} in Fig.~\ref{fig:composite}. 
At this junction, we need to enforce continuity of the tangential electric and magnetic fields. 
We can enforce the tangential magnetic field continuity by setting the two electric current coefficients $j_{2,n}$ and $j_{3,\tilde{n}}$ that are inside ${\cal V}_{2}$ and ${\cal V}_3$, respectively, to be equal to $j_{\mathrm{eq},n'}$ on ${\cal S}_{\mathrm{eq}}$ for some values of $n$, $n'$, and $\tilde{n}$. 
We can enforce this boundary condition by setting entries $(q^{j_{2,n}}, \tilde{q}^{j_{\mathrm{eq},n'}})$, $(q^{j_{\mathrm{eq},n'}}, \tilde{q}^{j_{\mathrm{eq},n'}})$ and $(q^{j_{3,\tilde{n}}}, \tilde{q}^{j_{\mathrm{eq},n'})}$ of $\matr{U}$ to $1$.
We can enforce continuity of the tangential electric field similarly.
\item {\bf PEC ground plane:} We consider the PEC ground plane interface shown in the region labeled with {\footnotesize \boxed{3}} in Fig.~\ref{fig:composite}. On this junction, the tangential electric field is zero.
The electric current densities on the two sides of the interface are independent. Therefore, two unique unknown coefficients are required to properly model this boundary condition. One of these unknowns is inside ${\cal S}_{\mathrm{eq}}$ and is collected in $\widetilde{\vect{X}}_{\mathrm{int}}$. The other is outside ${\cal S}_{\mathrm{eq}}$ and is, therefore, collected in $\widetilde{\vect{X}}_{\mathrm{eq}}$. To enforce this boundary condition, we set entries $({q}^{j_{1,n}}, \tilde{q}^{j_{1,n}})$ of $\matr{U}$ to $1$. 
We do not need an entry for $j_{\mathrm{eq},n'}$ since it does not appear in $\vect{X}$.

\item {\bf Junction at edges of a PEC ground plane:} Let us consider the junction at the edge of a PEC ground plane shown in the region labeled with {\footnotesize \boxed{4}} in Fig.~\ref{fig:composite}. The tangential electric field on this edge is zero.
Furthermore, the tangential magnetic fields on the two sides of the interface may or may not be independent depending on whether or not this unit cell is connected to other array unit cells also backed by a PEC ground plane. If the element is connected to another array element, then the current on two sides of the interface will be independent, leading to two unknowns. These two unknowns are collected into $\widetilde{\vect{X}}_{\mathrm{eq}}$.
We can implement this condition by setting entries $({q}^{j_{1,n}},\tilde{q}^{j_{1,n}})$ of $\matr{U}$ to $1$. If the element is not connected to another array element, then the tangential magnetic fields on two sides of the interface are equal. This condition will be enforced in Sec.~\ref{eq:ch5_macromodel_composite_bc} through another sparse matrix.
\end{enumerate}

By substituting~\eqref{eq:incidence1} into~\eqref{eq:discretized_SIE}, we obtain
\begin{equation}
\matr{Z} \matr{U} \widetilde{\vect{X}} = 0 \,,
\label{eq:mmm_mult}
\end{equation}
which is an over-determined system of equations. We eliminate additional equations by simply left-multiplying \eqref{eq:mmm_mult} by $\matr{U}^T$, to get
\begin{equation}
\matr{U}^T \matr{Z} \matr{U} \widetilde{\vect{X}} = 0 \,.
\label{eq:mmm_mult}
\end{equation}
This multiplication eliminates additional equations according to the PMCHWT formulation by adding the discretized EFIE and MFIE for the regions that share an interface or a junction~\cite{Gibson2009}. 

\subsection{Macromodel Generation}

Equation~\eqref{eq:mmm_mult} is a linear system of the form
\begin{equation}
\matr{U}^T \matr{Z} \matr{U} \widetilde{\vect{X}} = 
\begin{bmatrix}
\matr{Z}_{\mathrm{eq},\mathrm{eq}} & \matr{Z}_{\mathrm{eq}, \mathrm{int}}\\
\matr{Z}_{\mathrm{int},\mathrm{eq}} & \matr{Z}_{\mathrm{int},\mathrm{int}}
\end{bmatrix}
\begin{bmatrix}
\widetilde{\matr{X}}_{\mathrm{eq}}\\
\widetilde{\matr{X}}_{\mathrm{int}}
\end{bmatrix}
= 
\begin{bmatrix}
0 \\
0
\end{bmatrix}\,.
\end{equation}
Next, we eliminate $\widetilde{\vect{X}}_{\mathrm{int}}$ from our formulation by using the Schur complement and obtain
\begin{equation}
\underbrace{\left[ \matr{Z}_{\mathrm{eq},\mathrm{eq}}  - \matr{Z}_{\mathrm{eq}, \mathrm{int}} \matr{Z}_{\mathrm{int}, \mathrm{int}}^{-1} \matr{Z}_{\mathrm{int},\mathrm{eq}} \right]}_{\widetilde{\matr{Z}}_{\mathrm{eq},\mathrm{eq}}} \widetilde{\vect{X}}_{\mathrm{eq}} = 0\,.
\label{eq:Zmacromodel}
\end{equation}
Notice that after eliminating $\widetilde{\vect{X}}_{\mathrm{int}}$ from~\eqref{eq:Zmacromodel}, we have to solve for fewer unknowns that are associated with equivalent current densities on ${\cal S}_{\mathrm{eq}}$, instead of unknowns associated with current densities on PECs or interfaces of two dielectric regions. 
This is advantageous when simulating complex EM surfaces containing PEC traces with fine features. 
The proposed technique can lead to savings even when simulating an array of square patch antennas because in such structures patch antennas need to be meshed with very small triangular elements to resolve edge singularities in the current distribution~\cite{Zhou2015}.
Equation~\eqref{eq:Zmacromodel} describes the relation between the tangential electric and magnetic fields on ${\cal S}_{\mathrm{eq}}$, and can serve as a complete model for the electromagnetic behaviour of the objects inside ${\cal S}_{\mathrm{eq}}$, valid under any external excitation. Since this relation describes the behaviour of the unit cell using only field quantities defined on Seq,~\eqref{eq:Zmacromodel} can be interpreted as a \emph{macromodel} of the unit cell’s electromagnetic response.
As evident from~\eqref{eq:Zmacromodel}, generating the macromodel requires an LU factorization in order to eliminate interior unknowns. For a very complex unit cell, this step can be expensive. However, since we operate on a single unit cell, the relative complexity of this step is low compared to solving the entire array. 
Another advantage of the proposed macromodeling approach is that we only need to generate macromodels for unique elements. Therefore, even a complete electromagnetic surface with thousands of array elements may typically require the generation of comparatively few macromodels.
Furthermore, since $\widetilde{\matr{Z}}_{\mathrm{eq},\mathrm{eq}}$ depends only on geometrical and material properties of what is inside ${\cal S}_{\mathrm{eq}}$, the macromodel generation phase can be parallelized efficiently.

\section{Simulation of Electromagnetic Surfaces}
\label{sec:ch5_macromodel_composite_equivalent_exterior}

\subsection{Array of Scatterers}
We now consider the simulation of a large EM surface composed of $M$ unit cells.
Throughout the rest of this section, we will use superscript $(m)$ to denote the element number in the array. To simulate electromagnetic structures, we first create a macromodel for each unique element in the array. Then, we can express the relationship in~\eqref{eq:Zmacromodel} for all array elements as
\begin{equation}
\underbrace{
\begin{bmatrix}
\widetilde{\matr{Z}}_{\mathrm{eq},\mathrm{eq}}^{(1)} & 			& \\
								&   \ddots &  	\\
								& & \widetilde{\matr{Z}}_{\mathrm{eq},\mathrm{eq}}^{(M)}	
\end{bmatrix}}_{{\matr{Z}}_{\mathrm{eq}}}
\underbrace{
\begin{bmatrix}
\widetilde{\vect{X}}_{\mathrm{eq}}^{(1)} \\
\vdots \\
\widetilde{\vect{X}}_{\mathrm{eq}}^{(M)}
\end{bmatrix}}_{\matr{Y}}
= 
\begin{bmatrix} 
\vect{0} \\
\vdots \\
\vect{0}
\end{bmatrix}\,,
\label{eq:Zmacromodel_Array}
\end{equation}
where $\widetilde{\matr{Z}}_{\mathrm{eq},\mathrm{eq}}^{(m)}$ and $\widetilde{\vect{X}}^{(m)}_{\mathrm{eq}}$ are used to denote, respectively, the macromodel matrix and the list of current coefficients associated to ${\cal S}_{\mathrm{eq}}^{(m)}$, the equivalent surface for the $m$-th element.

\subsection{Inter-Element Coupling}

Next, we remove the scatterers inside ${\cal S}_{\mathrm{eq}}^{(m)}$ and model their presence through the equivalent electric current density 
$\vec{J}_{\mathrm{eq}}^{\,(m)}(\vr)$ and the equivalent magnetic current density $\vec{M}_{\mathrm{eq}}^{(m)}(\vr)$ introduced on ${\cal S}_{\mathrm{eq}}^{(m)}$ that radiate correct fields outside ${\cal S}_{\mathrm{eq}}^{(m)}$.
Furthermore, $\vec{J}_{\mathrm{eq}}^{\,(m)}(\vr)$ and $\vec{M}_{\mathrm{eq}}^{\,(m)}(\vr)$ on all ${\cal S}_{\mathrm{eq}}^{(m)}$ are related by the EFIE and MFIE
\begin{subequations}
\begin{align}
\hat{n} \times \vec{M}_{\mathrm{eq}}^{(m)}(\vr) =&\, \hat{n} \times \hat{n} \times \vec{E}^{\mathrm{inc}}(\vec{r}) \nonumber \\
& - j\omega\mu_0 \hat{n} \times \hat{n} \times  \left( \sum_{m'=1}^{M}\LArgtwo{\vec{J}_{\mathrm{eq}}^{\,(m')}(\vrp)}{o} \right) \nonumber \\ 
& + \hat{n} \times \hat{n} \times \left( \sum_{m'=1}^{M} \KArgtwo{\vec{M}_{\mathrm{eq}}^{(m')}(\vrp)}{o} \right) \label{eq:SIE_TEFIE1_Outer}\\
-\hat{n} \times \vec{J}_{\mathrm{eq}}^{\,(m)}(\vr) =&\, \hat{n} \times \hat{n} \times \vec{H}^{\mathrm{inc}}(\vec{r}) \nonumber \\
& + j\omega\varepsilon_0 \hat{n} \times \hat{n} \times \left( \sum_{m'=1}^{M} \LArgtwo{\vec{M}_{\mathrm{eq}}^{(m')}(\vrp)}{o} \right)  \nonumber \\
&- \hat{n} \times \hat{n} \times \left( \sum_{m'=1}^{M} \KArgtwo{\vec{J}_{\mathrm{eq}}^{\,(m')}(\vrp)}{o} \right)\,, \label{eq:SIE_TMFIE1_Outer}
\end{align}
\end{subequations}
where the $\vec{\cal L}_o$ and $\vec{\cal K}_o$ operators are computed with the material properties of free space (outer medium), and $\vec{E}^{\mathrm{inc}}$ and $\vec{H}^{\mathrm{inc}}$ are the incident electric and magnetic fields due to a feed antenna. 

We discretize the integral equations~\eqref{eq:SIE_TEFIE1_Outer}--\eqref{eq:SIE_TMFIE1_Outer} by substituting in them~\eqref{eq:Hdiscretebox}--\eqref{eq:Ediscretebox} and then testing them with RWG basis functions.
Discretized versions of~\eqref{eq:SIE_TEFIE1_Outer}--\eqref{eq:SIE_TMFIE1_Outer} can be compactly written as
\begin{align}
\underbrace{\begin{bmatrix}
\matr{Z}^{(1,1)}_{o} & \matr{Z}^{(1,2)}_{o} & \hdots & \matr{Z}^{(1,M)}_{o} \\
\matr{Z}^{(2,1)}_{o} & \matr{Z}^{(2,2)}_{o} & \hdots & \matr{Z}^{(2,M)}_{o} \\
\vdots & \vdots & \hdots & \vdots \\
\matr{Z}^{(M,1)}_{o} & \matr{Z}^{(M,2)}_{o} & \hdots & \matr{Z}^{(M,M)}_{o} \\
\end{bmatrix}}_{\matr{Z}_{o}}
\begin{bmatrix}
\widetilde{\vect{X}}_{\mathrm{eq}}^{(1)} \\
\widetilde{\vect{X}}_{\mathrm{eq}}^{(2)} \\
\vdots \\
\widetilde{\vect{X}}_{\mathrm{eq}}^{(M)}
\end{bmatrix}
=
\underbrace{\begin{bmatrix}
\vect{V}^{(1)} \\
\vect{V}^{(2)} \\
\vdots \\
\vect{V}^{(M)}
\end{bmatrix}}_{\vect{V}} \,,
\label{eq:out}
\end{align}
where 
\begin{equation}
\matr{Z}^{(m,m')}_{o} = \begin{bmatrix} \matr{L}_o^{E, (m,m')} & \matr{K}_o^{E, (m,m')} \\ \matr{K}_o^{H,(m,m')} & \matr{L}_o^{H,(m,m')}\end{bmatrix}
\label{eq:blockZo}
\end{equation}
stores the discretized~\eqref{eq:SIE_TEFIE1_Outer}--\eqref{eq:SIE_TMFIE1_Outer} when source basis functions are on ${\cal S}_{\mathrm{eq}}^{(m')}$ and test basis functions are on ${\cal S}_{\mathrm{eq}}^{(m)}$. Matrix $\matr{Z}^{(m,m')}_{o}$ captures the mutual coupling between the macromodels of the $m$-th and $m'$-th elements of the array. 
In~\eqref{eq:out}, $\vect{V}^{(m)}$ is the excitation vector that is generated by testing the incident electric and magnetic fields with RWG basis functions on ${\cal S}_{\mathrm{eq}}^{(m)}$.

\subsection{Boundary Conditions}
\label{eq:ch5_macromodel_composite_bc}
Now, we have two sets of equations. 
The first set of equations is the macromodel equation~\eqref{eq:Zmacromodel_Array}, which captures the electromagnetic behavior of the scatterers inside each ${\cal S}_{\mathrm{eq}}^{(m)}$ for $m = 1,\hdots, M$. The second set of equations is the discretized EFIE and MFIE for the outer problem~\eqref{eq:out}, which captures the mutual coupling between the macromodels.
Hence, we have more equations than unknowns. 
Therefore, as in the PMCHWT formulation, we will add up~\eqref{eq:Zmacromodel_Array} and~\eqref{eq:out} to form a full-rank, well-conditioned system of linear equations of the form
\begin{align}
\left( \matr{Z}_{\mathrm{eq}} + \matr{Z}_{o}\right) \vect{Y} = \vect{V}\,.
\label{eq:out3}
\end{align}

\subsection{Enforcing Boundary Conditions on Fictitious Surfaces}
\label{sec:ch5_macromodel_composite_ground_plane}

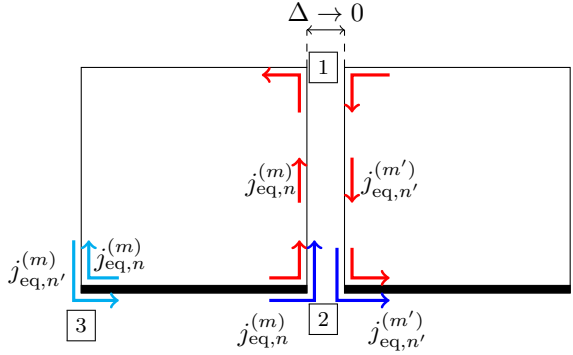
\begin{figure}[t]
\centering
\begin{tikzpicture}

\draw (0,0) rectangle (3, 3);
\draw[fill=black] (0,0) rectangle (3,0.1);
\draw[red, very thick, ->] (2.9,1.2) -- (2.9,1.8);
\draw[red, very thick, ->] (2.9,2.4) -- (2.9,2.9) -- (2.4,2.9);
\draw[red, very thick, ->] (2.5,0.2) -- (2.9,0.2) -- (2.9,0.7);
\draw[blue, very thick, ->] (2.5,-0.1) -- (3.1,-0.1) -- (3.1,0.7);
\node at (0.05,0.5) [right] {$j_{\mathrm{eq},n}^{(m)}$};
\node at (-0.05,0.3) [left] {$j_{\mathrm{eq},n'}^{(m)}$};

\node at (3,1.5) [left] {$j_{\mathrm{eq},n}^{(m)}$};

\draw[cyan, very thick, ->] (0.5,0.2) -- (0.1,0.2) -- (0.1,0.7);
\draw[cyan, very thick, <-] (0.5,-0.1) -- (-0.1,-0.1) -- (-0.1,0.7);
\node at (0,-0.1) [below]{\footnotesize \boxed{3}};
\node at (2.9,3.0) [right]{\footnotesize \boxed{1}};
\node at (2.9,-0.35) [right]{\footnotesize \boxed{2}};
\node at (2.9,-0.5) [left] {$j_{\mathrm{eq},n}^{(m)}$};
\node at (3.25,3.5) [above] {$\Delta \rightarrow 0$};
\draw[dashed] (3,3) -- (3,3.5);
\draw[dashed] (3.5,3) -- (3.5,3.5);
\draw[<->] (3.0,3.5) -- (3.5,3.5);
\begin{scope}[shift = {(0.5,0)}]
\draw (3,0) rectangle (6, 3);
\draw[fill=black] (3,0) rectangle (6,0.1);
\draw[red, very thick, <-] (3.1,1.2) -- (3.1,1.8);
\draw[red, very thick, <-] (3.1,2.4) -- (3.1,2.9) -- (3.6,2.9);
\draw[red, very thick, <-] (3.6,0.2) -- (3.1,0.2) -- (3.1,0.6);
\draw[blue, very thick, <-] (3.6,-0.1) -- (2.9,-0.1) -- (2.9,0.6);
\node at (3.1,1.5) [right] {$j_{\mathrm{eq},n'}^{(m')}$};
\node at (3.2,-0.5) [right] {$j_{\mathrm{eq},n'}^{(m')}$};

\end{scope}



\end{tikzpicture}
\caption{A sample array of two fictitious surfaces that are touching one another. In the graphics, we have shown some space between the boxes to draw current directions. Three boundary conditions that need to be enforced when equivalent surfaces are connected are shown in red, blue, and cyan.} 
\label{fig:connections}
\end{figure}

When simulating planar electromagnetic structures, ${\cal S}_{\mathrm{eq}}^{(m)}$ partially overlaps with ${\cal S}_{\mathrm{eq}}^{(m')}$ if the $m'$-th unit cell is adjacent to the $m$-th unit cell.
Therefore, tangential electric and magnetic current densities on the overlapping surface may have been expanded with a duplicated set of basis functions, resulting in redundant unknowns.
We eliminate redundant unknowns and enforce proper boundary conditions through another sparse matrix $\matr{U}_{o}$. 
Matrix $\matr{U}_o$ relates $\vect{Y}$ to a vector of unique unknowns $\widetilde{\vect{Y}}$ through
\begin{equation}
\vect{Y} = \matr{U}_o \widetilde{\vect{Y}}\,.
\label{eq:connections_IUI}
\end{equation}
To discuss boundary conditions, we consider a sample array with two fictitious surfaces, as shown in Fig.~\ref{fig:connections}.
Matrix $\matr{U}_o$ is generated by applying the following boundary conditions:
\begin{enumerate}
\item {\bf Surface common to two equivalent surfaces:} The tangential electric and magnetic fields are equal on the overlapping surface between  ${\cal S}_{\mathrm{eq}}^{(m)}$ and ${\cal S}_{\mathrm{eq}}^{(m')}$. These fields are  depicted in red and labeled with {\footnotesize \boxed{1}} in Fig.~\ref{fig:connections}. To enforce this boundary condition, the electric current density coefficient $j_{\mathrm{eq},n}^{(m)}$ on ${\cal S}_{\mathrm{eq}}^{(m)}$ is set to be equal to $j_{\mathrm{eq},n'}^{(m')}$ on ${\cal S}_{\mathrm{eq}}^{(m')}$. Continuity of the tangential magnetic field  is enforced similarly.
\item {\bf Intersection of PEC ground planes of adjacent unit cells:} The electric current densities on the two sides of a ground plane on the edge of a unit cell are independent. Therefore, when ${\cal S}_{\mathrm{eq}}^{(m)}$ and ${\cal S}_{\mathrm{eq}}^{(m')}$ are connected, we need to correctly enforce continuity of the electric current density on both sides of the ground plane. Hence, on both sides of the ground plane, we equate the electric current density coefficients $j_{\mathrm{eq},n}^{(m)}$ on ${\cal S}_{\mathrm{eq}}^{(m)}$ and  $j_{\mathrm{eq},n'}^{(m')}$ on ${\cal S}_{\mathrm{eq}}^{(m')}$, as shown in the region labeled with {\footnotesize \boxed{2}} in Fig.~\ref{fig:connections}.  
\item {\bf Edges of a PEC ground plane:} We consider the edge of a PEC ground plane shown in the region labeled with {\footnotesize \boxed{3}} in Fig.~\ref{fig:connections}. On this edge, the electric current coefficients $j_{\mathrm{eq},n}^{(m)}$ and $j_{\mathrm{eq},n'}^{(m)}$ on the two sides of the ground plane are set to be equal in order to satisfy the continuity of the tangential magnetic field.
\end{enumerate}

We substitute~\eqref{eq:connections_IUI} into~\eqref{eq:out3}, and left-multiply the resulting equation by $\matr{U}_o^T$ to eliminate additional equations as in the PMCHWT formulation. The final equation is given by
\begin{align}
\matr{U}_{\mathrm{o}}^T \left[ \matr{Z}_{\mathrm{eq}} + \matr{Z}_{o} \right] \matr{U}_{\mathrm{o}} \widetilde{\vect{Y}} &= \underbrace{\matr{U}_{\mathrm{o}}^T \vect{V}}_{\widetilde{\matr{V}}}\,,
\label{eq:final}
\end{align}
where $\widetilde{\matr{V}}$ is the excitation vector obtained after enforcing boundary conditions. We can solve~\eqref{eq:final} using a direct or an iterative solver to obtain current density coefficients on ${\cal S}_{\mathrm{eq}}^{(m)}$ for $m = 1,\hdots, M$. Once we have computed the tangential electric and magnetic fields on the equivalent surface of each unit cell, we can compute fields scattered from the electromagnetic structure through the EFIE and MFIE in conjunction with the free space Green's function.
Note that the final set of unknowns in~\eqref{eq:final} corresponds to unknowns that are only on ${\cal S}_{\mathrm{eq}}^{(m)}$, and not on the scatterers inside ${\cal S}_{\mathrm{eq}}^{(m)}$.

\section{Accelerating Matrix-Vector Products with the FFT}
\label{sec:ch5_macromodel_composite_toeplitz}

\subsection{Iterative solver}

When simulating large EM surfaces with many unit cells, we solve~\eqref{eq:final} iteratively with GMRES~\cite{petsc-user-ref}.
For this step, we need an efficient preconditioner and a fast technique to compute matrix-vector products.

In our formulation, we use the preconditioner matrix
\begin{equation}
\matr{P} = \matr{U}_{o}^T \left[ \matr{Z}_{\mathrm{eq}}^{\mathrm{NF}} + \matr{Z}_{o}^{\mathrm{NF}} \right] \matr{U}_{o} \,,
\end{equation}
where $\matr{Z}_{\mathrm{eq}}^{\mathrm{NF}}$ and $\matr{Z}_{o}^{\mathrm{NF}}$ collect near-field entries of $\matr{Z}_{\mathrm{eq}}$ and $\matr{Z}_{o}$, respectively. That is, the $(p,q)$-th entry of $\matr{Z}_{\mathrm{eq}}^{\mathrm{NF}}$ and $\matr{Z}_{o}^{\mathrm{NF}}$ is non-zero, and is equal to the $(p,q)$-th entry of $\matr{Z}_{\mathrm{eq}}$ and $\matr{Z}_{o}$, respectively, if the distance between basis functions associated with the $p$-th and $q$-th unknowns is less than $\Delta_{\mathrm{NF}}$. In our simulations, we use $\Delta_{\mathrm{NF}}$ to be between $\lambda_0/10$ and $\lambda_0/6$ depending on the periodicity of the array. We apply $\matr{P}$ as a right preconditioner, obtaining
\begin{equation}
\matr{U}_{\mathrm{o}}^T \left[ \matr{Z}_{\mathrm{eq}} + \matr{Z}_o \right] \matr{U}_{\mathrm{o}} \matr{P}^{-1} \widetilde{\vect{Y}}'  = \widetilde{\vect{V}} \,,
\label{eq:final_precond}
\end{equation}
for $\widetilde{\vect{Y}}' = \matr{P} \widetilde{\vect{Y}}$.
To solve~\eqref{eq:final_precond}, we need to evaluate $\matr{P}^{-1} \vect{x}$, given some vector $\vect{x}$. Since $\matr{P}$ is  sparse, we use an LU factorization to compute $\matr{P}^{-1} \vect{x}$.

An iterative solver also requires the computation of
\begin{equation}
\matr{U}_{\mathrm{o}}^T \left[ \matr{Z}_{\mathrm{eq}} + \matr{Z}_{o} \right] \matr{U}_{\mathrm{o}} \vect{x} \,,
\label{eq:MVP}
\end{equation}
given $\vect{x}$. 
In this matrix-vector multiplication, $\vect{x}_1 = \matr{U}_{o} \vect{x}$ is very cheap to compute because $\matr{U}_{o}$ is very sparse. Furthermore, since $\matr{Z}_{\mathrm{eq}}$ is block diagonal, the matrix-vector product $\matr{Z}_{\mathrm{eq}} \vect{x}_1$ is also inexpensive. 
However, $\matr{Z}_o \vect{x}_1$ is expensive to compute because $\matr{Z}_o$ is dense. Furthermore, storing $\matr{Z}_o$ in a dense format explicitly is not feasible.
Therefore, we need to apply an acceleration algorithm to compute $\matr{Z}_o\vect{x}_1$.
In most integral equation methods, this matrix-vector product is accelerated with the MLFMM~\cite{MLFMM} or AIM~\cite{Bleszynski96}.
However, we exploit two properties of the problem at hand to accelerate the computation of $\matr{Z}_o \vect{x}_1$ via FFTs. First, the equivalent electric and magnetic current densities on each ${\cal S}_{\mathrm{eq}}^{(m)}$ are expanded using identical sets of RWG basis functions because ${\cal S}_{\mathrm{eq}}^{(m)}$ are meshed identically. Second, electromagnetic structures have constant periodicity along transverse directions. Since the Green's function of the free space medium is translation-invariant, these two properties combine to give $\matr{Z}_o$ with a Toeplitz form~\cite{Gray2006}.

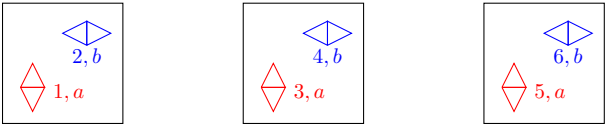
\begin{figure}[t]
\centering
\begin{tikzpicture}[scale=0.8, every node/.style={scale=0.8}]

\draw (0,0) rectangle (2, 2);
\draw[red] (0.3,0.6) -- (0.5,1) -- (0.7,0.6) -- (0.3,0.6);
\draw[red] (0.3,0.6) -- (0.5,0.2) -- (0.7,0.6) -- (0.3,0.6);
\node at (1.1,0.5) {\color{red}$1,a$};
\node at (1.4,1.1) {\color{blue}$2,b$};
\draw[blue] (1.4,1.3) -- (1.8,1.5) -- (1.4,1.7) -- (1.4,1.3);
\draw[blue] (1.4,1.3) -- (1.0,1.5) -- (1.4,1.7) -- (1.4,1.3);

\begin{scope}[shift = {(4,0)}]
\draw (0,0) rectangle (2, 2);
\draw[red] (0.3,0.6) -- (0.5,1) -- (0.7,0.6) -- (0.3,0.6);
\draw[red] (0.3,0.6) -- (0.5,0.2) -- (0.7,0.6) -- (0.3,0.6);
\node at (1.1,0.5) {\color{red}$3,a$};
\node at (1.4,1.1) {\color{blue}$4,b$};
\draw[blue] (1.4,1.3) -- (1.8,1.5) -- (1.4,1.7) -- (1.4,1.3);
\draw[blue] (1.4,1.3) -- (1.0,1.5) -- (1.4,1.7) -- (1.4,1.3);
\end{scope}

\begin{scope}[shift = {(8,0)}]
\draw (0,0) rectangle (2, 2);
\draw[red] (0.3,0.6) -- (0.5,1) -- (0.7,0.6) -- (0.3,0.6);
\draw[red] (0.3,0.6) -- (0.5,0.2) -- (0.7,0.6) -- (0.3,0.6);
\node at (1.1,0.5) {\color{red}$5,a$};
\node at (1.4,1.1) {\color{blue}$6,b$};
\draw[blue] (1.4,1.3) -- (1.8,1.5) -- (1.4,1.7) -- (1.4,1.3);
\draw[blue] (1.4,1.3) -- (1.0,1.5) -- (1.4,1.7) -- (1.4,1.3);
\end{scope}

\end{tikzpicture}
\caption{Top view of a sample array of three fictitious surfaces with identical meshes. Each surface has two identical basis functions.
Each basis function has a local and a global identification number.
Local identification numbers are denoted by $a$ and $b$. Global identification numbers are denoted by $1, \hdots, 6$.}
\label{fig:ch5_macromodel_composite_toeplitz}
\end{figure}

\subsection{Evaluation of Matrix-Vector Product with the FFT}
\label{sec:ch5_macromodel_composite_toeplitz_MV}

To discuss how to accelerate the computation of $\matr{Z}_o \vect{x}$ with the fast Fourier transform~\cite{FFTW05}, we consider the scenario in Fig.~\ref{fig:ch5_macromodel_composite_toeplitz}. 
This array has three fictitious surfaces that are uniformly spaced and have identical meshes.
For the sake of simplicity, let us consider that electric and magnetic current densities on each surface are expanded with only two basis functions.
Each basis function in the array is assigned a local and a global identification number. Local identification numbers are denoted by $a$ and $b$. Global identification numbers are denoted by $1, \hdots, 6$.

We can see from~\eqref{eq:blockZo} that $\matr{Z}_o$ is generated by discretizing the $\vec{\cal L}_o$ and $\vec{\cal K}_o$ operators. Therefore, $\matr{Z}_o \vect{x}$ can be evaluated by multiplying the discretized $\vec{\cal L}_o$ and $\vec{\cal K}_o$ operators, scaled by appropriate constants, with a block of column vector $\vect{x}$. Since equivalent surfaces have identical meshes  and are uniformly spaced, and the free space Green's function is translation-invariant, both the discretized $\vec{\cal L}_o$ and $\vec{\cal K}_o$ operators can be cast into Toeplitz matrices~\cite{Gray2006}.
For the example shown in Fig.~\ref{fig:ch5_macromodel_composite_toeplitz}, the matrix-vector product with the discretized $\vec{\cal L}_o$ operator can be written as
\begin{equation}
\begin{bmatrix}
{{L}_{11}^{aa}} & {{L}_{12}^{ab}} & {{L}_{13}^{aa}} & {{L}_{14}^{ab}} &  {{L}_{15}^{aa}} & {{L}_{16}^{ab}}\\
{{L}_{21}^{ba}} & {L}_{22}^{bb} & {{L}_{23}^{ba}} & {L}_{24}^{bb} & {{L}_{25}^{ba}} & {L}_{26}^{bb} \\
{{L}_{31}^{aa}} & {{L}_{32}^{ab}} & {{L}_{33}^{aa}} & {{L}_{34}^{ab}} & {{L}_{35}^{aa}} & {{L}_{36}^{ab}} \\
{{L}_{41}^{ba}} & {L}_{42}^{bb} & {{L}_{43}^{ba}} & {L}_{44}^{bb} & {{L}_{45}^{ba}} & {L}_{46}^{bb} \\  
{{L}_{51}^{aa}} & {{L}_{52}^{ab}} & {{L}_{53}^{aa}} & {{L}_{54}^{ab}} & {{L}_{55}^{aa}} & {{L}_{56}^{ab}} \\
{{L}_{61}^{ba}} & {L}_{62}^{bb} & {{L}_{63}^{ba}} & {L}_{64}^{bb} & {{L}_{65}^{ba}} & {L}_{66}^{bb} \\
\end{bmatrix}
\begin{bmatrix}
J_{1}^{a} \\
J_{2}^{b} \\
J_{3}^a \\
J_4^b \\
J_5^a \\
J_6^b \\
\end{bmatrix} \,,
\end{equation}
where ${L}_{mn}^{ij}$ denotes the reaction term due to the $n$-th source basis function ($j$-th local basis function) and the $m$-th test basis function  ($i$-th local basis function). Similarly, $J_n^i$ is the $n$-th source coefficient.
 
Now let us rearrange the matrix and the excitation vector by grouping together identical basis functions on the equivalent surfaces. By doing this, we obtain
\begin{align}
\left[
\begin{array}{ccc|ccc}
{{L}_{11}^{aa}} & {{L}_{13}^{aa}} & {{L}_{15}^{aa}} & {{L}_{12}^{ab}} &  {{L}_{14}^{ab}} &  {{L}_{16}^{ab}} \\
{{L}_{31}^{aa}} & {{L}_{33}^{aa}} & {{L}_{35}^{aa}} & {{L}_{32}^{ab}}  & {{L}_{34}^{ab}} & {{L}_{36}^{ab}} \\ 
{{L}_{51}^{aa}} & {{L}_{53}^{aa}} & {{L}_{55}^{aa}} & {{L}_{52}^{ab}}  & {{L}_{54}^{ab}} & {{L}_{56}^{ab}} \\ \hline
{{L}_{21}^{ba}} & {{L}_{23}^{ba}} & {{L}_{25}^{ba}} & {L}_{22}^{bb}  & {L}_{24}^{bb} & {L}_{26}^{bb} \\
{{L}_{41}^{ba}} & {{L}_{43}^{ba}} & {{L}_{45}^{ba}} & {L}_{42}^{bb} &  {L}_{44}^{bb} & {L}_{46}^{bb} \\
{{L}_{61}^{ba}} & {{L}_{63}^{ba}} & {{L}_{65}^{ba}} & {L}_{62}^{bb} &  {L}_{64}^{bb} & {L}_{66}^{bb} \\
\end{array}
\right]
\left[
\begin{array}{c}
J_{1}^{a} \\
J_{3}^a \\ 
J_{5}^a  \\ \hline
J_{2}^{b} \\
J_4^b \\
J_6^b 
\end{array} \right] \,,
\label{eq:ch5_macromodel_composite_LJV1}
\end{align}
where the matrix is subdivided into 4 blocks, each of which collects reaction integrals between a pair of local basis functions.
We can compactly write~\eqref{eq:ch5_macromodel_composite_LJV1} as
\begin{align}
\left[
\begin{array}{cc}
\matr{L}^{aa} & \matr{L}^{ab}\\
\matr{L}^{ba} & \matr{L}^{bb} 
\end{array}
\right]
\left[
\begin{array}{c}
\vect{J}^{a} \\
\vect{J}^{b}
\end{array} \right] \,,
\label{eq:ch5_macromodel_composite_LJV}
\end{align}
where $\matr{L}^{aa}$, $\matr{L}^{ab}$, $\matr{L}^{ba}$, and $\matr{L}^{bb}$ are $3\times 3$ Toeplitz matrices~\cite{Gray2006}, and thus we can use FFTs to evaluate their products with a vector. 
Here, we demonstrate how to use FFT to compute $\matr{L}^{aa} \vect{J}^a$. Other matrix-vector products can be computed similarly. To compute $\matr{L}^{aa} \vect{J}^a$, we augment $\matr{L}^{aa}$ to form a circulant matrix~\cite{Gray2006}
\begin{equation}
\matr{L}_{aa}' = 
\begin{bmatrix}
\red{{L}^{aa}_{11}} & \blue{{L}^{aa}_{13}} & \green{{L}^{aa}_{15}} & \brown{{L}^{aa}_{51}} & {L}^{aa}_{31}\\
{L}^{aa}_{31} & \red{{L}^{aa}_{33}} & \blue{{L}^{aa}_{35}} & \green{{L}^{aa}_{15}} & \brown{L^{aa}_{51}} \\
\brown{{L}^{aa}_{51}} & {L}^{aa}_{53} & \red{{L}^{aa}_{55}} & \blue{L^{aa}_{35}} & \green{L^{aa}_{15}} \\
\green{L^{aa}_{15}} & \brown{L^{aa}_{51}} & L^{aa}_{53} & \red{L^{aa}_{55}} & \blue{L^{aa}_{35}} \\
\blue{L^{aa}_{13}} & \green{L^{aa}_{15}} & \brown{L^{aa}_{51}} & L^{aa}_{53} & \red{L^{aa}_{55}}
\end{bmatrix}\,,
\label{eq:ch5_macromodel_composite_LJV3}
\end{equation}
where we have color-coded the entries that have the same value.
Using the circulant matrix $\matr{L}_{aa}'$, the matrix-vector product is~\cite{Gray2006}
\begin{equation}
\begin{bmatrix} \matr{L}^{aa} \vect{J}^a \\ * \\ *  \end{bmatrix} = \begin{bmatrix}
\red{{L}^{aa}_{11}} & \blue{{L}^{aa}_{13}} & \green{{L}^{aa}_{15}} & \brown{{L}^{aa}_{51}} & {L}^{aa}_{31}\\
{L}^{aa}_{31} & \red{{L}^{aa}_{33}} & \blue{{L}^{aa}_{35}} & \green{{L}^{aa}_{15}} & \brown{L^{aa}_{51}} \\
\brown{{L}^{aa}_{51}} & {L}^{aa}_{53} & \red{{L}^{aa}_{55}} & \blue{L^{aa}_{35}} & \green{L^{aa}_{15}} \\
\green{L^{aa}_{15}} & \brown{L^{aa}_{51}} & L^{aa}_{53} & \red{L^{aa}_{55}} & \blue{L^{aa}_{35}} \\
L^{aa}_{13} & \green{L^{aa}_{15}} & \brown{L^{aa}_{51}} & L^{aa}_{53} & \red{L^{aa}_{55}}
\end{bmatrix} \begin{bmatrix} J_1^a \\ J_3^a \\ J_5^a \\ 0 \\ 0 \end{bmatrix} \,,
\label{eq:ch5_macromodel_composite_LJV4}
\end{equation}
where $*$ denotes values that are not useful to us.
The right-hand side of~\eqref{eq:ch5_macromodel_composite_LJV4} can be calculated as
\begin{equation}
\begin{bmatrix} \matr{L}^{aa} \vect{J}^a \\  * \\ * \end{bmatrix}  =  {\cal F}^{-1} \left[ \widetilde{\matr{L}}^{aa} \cdot \widetilde{\vect{J}}^{a} \right]\,,
\label{eq:ch5_macromodel_composite_LJV5}
\end{equation}
where ${\cal F}^{-1}[\cdot]$ is the inverse fast Fourier transform operator~\cite{FFTW05}, ``$\cdot$'' denotes element-wise multiplication, and
\begin{subequations}
\begin{align}
\widetilde{\matr{L}}^{aa} &=  {\cal F} \begin{bmatrix} \red{{L}^{aa}_{11}} & \blue{{L}^{aa}_{13}} & \green{{L}^{aa}_{15}} & \brown{{L}^{aa}_{51}} & {L}^{aa}_{31} \end{bmatrix}^T \,, \\
\widetilde{\vect{J}}^{a} &= {\cal F} \begin{bmatrix}  J_1^a & J_3^a & J_5^a & 0 & 0 \end{bmatrix}^T \,,
\end{align}
\end{subequations}
where ${\cal F}[\cdot]$ is the fast Fourier transform operator~\cite{FFTW05}.
The matrix-vector products involving the discretized $\vec{\cal K}_o$ operator can be accelerated using a similar procedure.
Hence, the matrix-vector product in~\eqref{eq:ch5_macromodel_composite_LJV} requires performing four 1-D FFT operations. 
Note that, while we only demonstrated the matrix-vector product acceleration for a 1-D array, the technique can be generalized to 2-D and 3-D arrays using higher-dimensional FFTs.

\subsection{Discussion}
Recall that, in our formulation, there are $M$ equivalent surfaces, and the tangential electric and magnetic fields on each surface are discretized with at most $N_{\mathrm{eq}}$ RWG basis functions\footnote{The number of basis functions will be less than $N_{\mathrm{eq}}$ if some portions of ${\cal S}_{\mathrm{eq}}$ extend over a PEC ground plane.}.
Therefore, the computational cost of evaluating the matrix-vector product with the proposed Toeplitz method is ${\cal O}\left(N_{\mathrm{eq}}^2 M \log_2 M \right)$. 
Furthermore, the proposed approach requires storing only $2^{d+1} N_{\mathrm{eq}}^2M$ complex numbers for a $d$-dimensional array.

The Toeplitz acceleration method is simpler to implement than the MLFMM~\cite{MLFMM} and AIM~\cite{Zhu2005} because it avoids a lot of overhead costs associated with the MLFMM~\cite{MLFMM} and AIM~\cite{Zhu2005}. For example, the AIM requires computing and storing projection and interpolation matrices, which are not needed  with the proposed method. The AIM also requires pre-correction, which is not necessary with the proposed technique. Likewise, the MLFMM also requires aggregation and deaggregation steps, which are not needed in the proposed technique. 
Furthermore, the Toeplitz acceleration method can also be parallelized more efficiently than MLFMM or AIM due to low overhead cost.
Previously, the Toeplitz acceleration technique was applied to analyze arrays of identical scatterers with non-zero spacing between adjacent elements~\cite{Kin03}.
The proposed work generalizes this approach to uniform arrays of dissimilar scatterers, allowing us to exploit the computational savings offered by a Toeplitz structure even when scatterers are not the same. 
This possibility is one of the most promising features of the proposed macromodeling approach, that leads to a ``periodic’’ distribution of unknowns even for structures that are not periodic.

\section{Numerical Results}
\label{sec:Results}
We consider the simulation of two reflectarrays to validate the proposed macromodeling technique. 
\subsection{A Single-Layer Reflectarray with Square Patches}
\label{sec:SingleLayerReflectarray}

\begin{figure}[t]
\begin{center}
\null \hfill
\subfloat[\label{fig:cross_section_16by16} $16 \times 16$ Array] {
\includegraphics[width = 0.46\columnwidth]{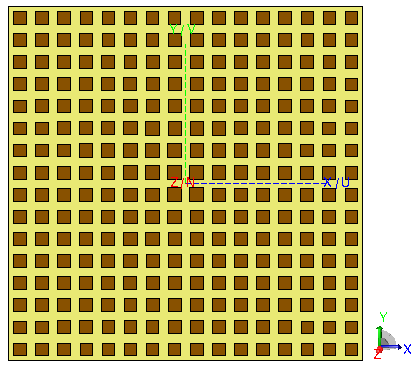}} \hfill
\subfloat[\label{fig:cross_section_30by30} $30 \times 30$ Array] {
\includegraphics[width = 0.46\columnwidth]{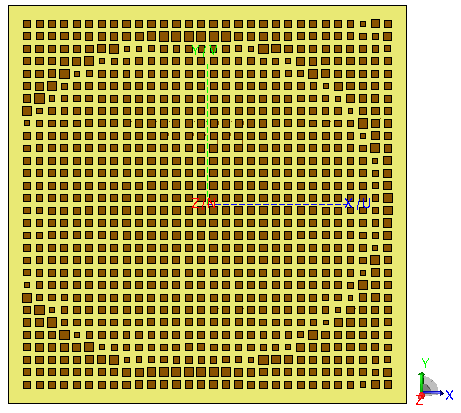}} \hfill \null
\end{center}
\caption{Top view of reflectarrays considered in Sec.~\ref{sec:SingleLayerReflectarray}}
\label{fig:Topview_squarepatch}
\end{figure}

We first consider a single-layer reflectarray comprised of square patch elements. 
This structure was previously presented in~\cite{Zhou2015,APS2019}. 
The reflectarray dielectric substrate is backed by a PEC ground plane and has a relative permittivity $\varepsilon_r = 3.66$ and a thickness of $0.762~\mathrm{mm}$. 
Each unit cell of the reflectarray is $13.5~\mathrm{mm} \times 13.5~\mathrm{mm}$.
The width of each square patch varies between $5.4~\mathrm{mm}$ and $10~\mathrm{mm}$.
The reflectarray is placed in the $xy$ plane and is centered about the $z$-axis.
It is excited by a linearly-polarized corrugated horn antenna operating at $f = 9.6~\mathrm{GHz}$, which we modeled with a spherical wave expansion derived from a measured horn antenna~\cite{Zhou2015}.
The horn antenna is centered at $(0.30~\mathrm{m}, 0, 0.52~\mathrm{m})$ and points towards the center of the reflectarray, i.e. the horn antenna is $30^\circ$ off the axis of the reflectarray.
The reflectarray is designed to collimate the main beam in the $(\phi = 180^\circ, \theta = 30^{\circ})$ direction. 
We simulated two different sizes of this reflectarray: $16 \times 16$ and $30 \times 30$. 
The $16 \times 16$ reflectarray is a subset of the $30 \times 30$ array, formed by the central 256 elements of the $30 \times 30$ reflectarray.

\begin{figure}[t]
\begin{center}
\null \hfill
\subfloat[\label{fig:reflectarray_16by16_cut1} $\phi = 0^{\circ}$] {
\includegraphics[width=0.95\columnwidth]{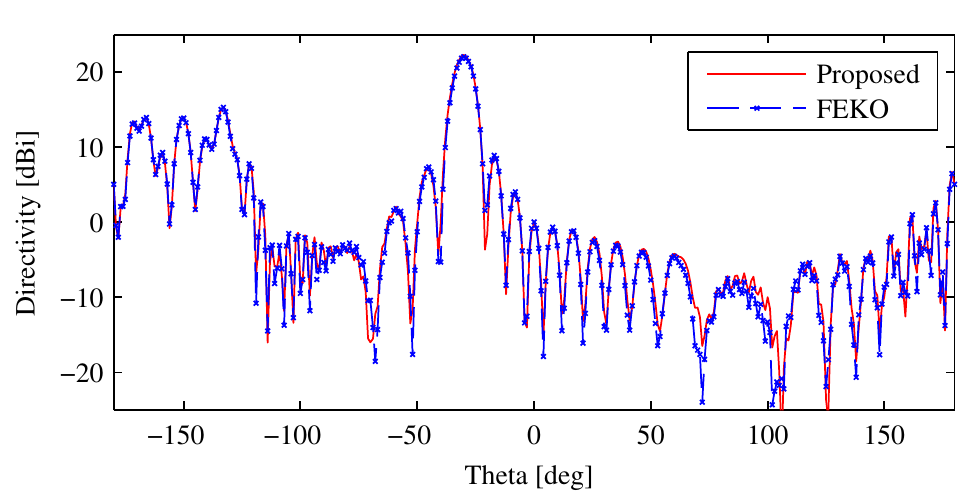} 
}
\hfill \null \\
\null \hfill
\subfloat[\label{fig:reflectarray_16by16_cut2} $\phi = 90^{\circ}$] {
\includegraphics[width=0.95\columnwidth]{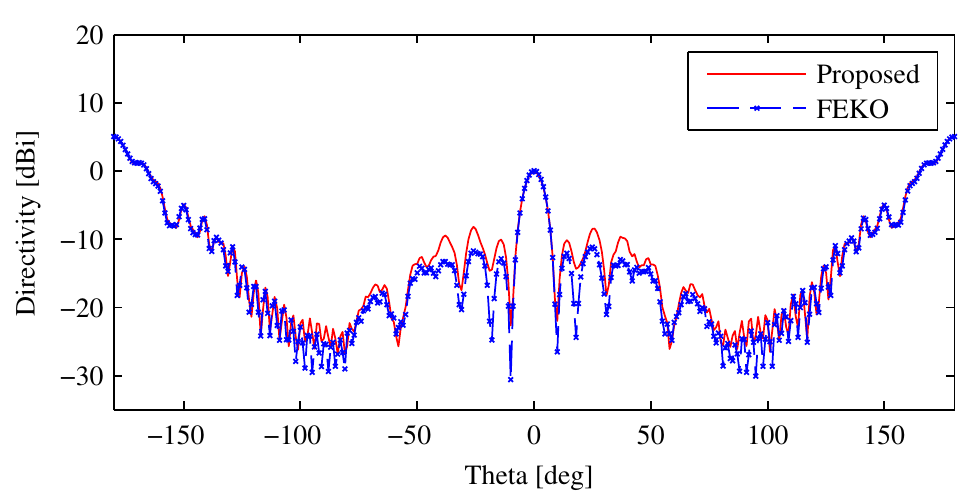}} \hfill \null \\
\null \hfill 
\subfloat[\label{fig:reflectarray_16by16_cut3} $\phi = 45^{\circ}$] {
\includegraphics[width=0.95\columnwidth]{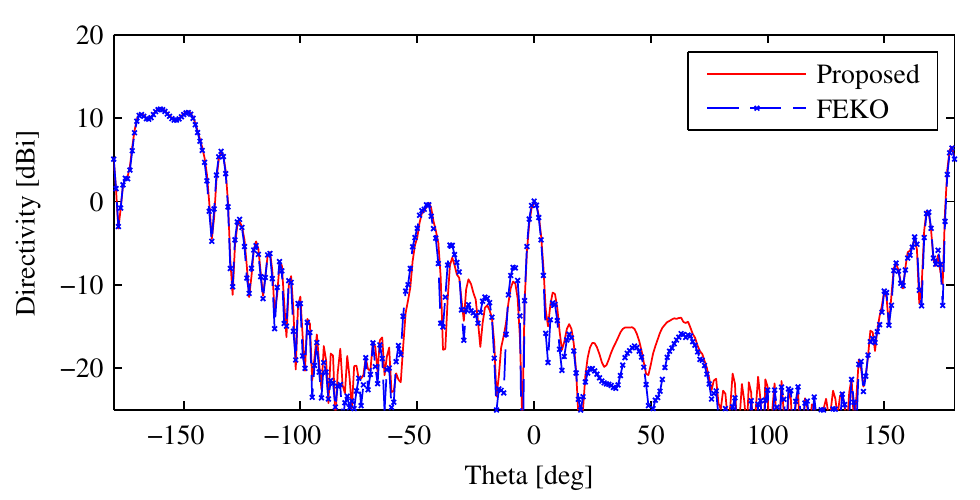}}
\hfill \null
\end{center}
\caption{Directivity of the $16 \times 16$ reflectarray considered in Sec.~\ref{sec:ticra_16by16} calculated with FEKO and with the proposed technique.}
\label{fig:directivity_16by16}
\end{figure}

\subsubsection{${16 \times 16}$ Element Reflectarray}
\label{sec:ticra_16by16}

This 256-element array features nine distinct square patch sizes. A top view of this reflectarray is shown in Fig.~\ref{fig:cross_section_16by16}. 
We simulated this structure with the MLFMM in FEKO~\cite{FEKO} and with the proposed macromodeling technique.
In both simulations, we meshed patch antennas with very small triangular elements with a characterstic length of $0.80~\mathrm{mm}$ in order to accurately resolve edge singularities in current density. 
The characteristic mesh length along the dielectric substrate was chosen to be $1.75~\mathrm{mm}$ in both techniques.
In the macromodeling approach, each unit cell was enclosed by a fictitious surface ${\cal S}_{\mathrm{eq}}^{(m)}$ of size $13.5~\mathrm{mm} \times 13.5~\mathrm{mm} \times 2~\mathrm{mm}$, with the top region set to have material properties of free space. 
The bottom surface of ${\cal S}_{\mathrm{eq}}^{(m)}$ coincided with the PEC ground plane.
The equivalent surface was discretized with a characteristic mesh length of $2.5~\mathrm{mm}$.

Figure~\ref{fig:directivity_16by16} shows the directivity of the reflectarray in the $\phi = 0^{\circ}$, $\phi = 45^{\circ}$, and $\phi = 90^{\circ}$ cuts. Results obtained with the macromodeling approach and FEKO match very well, validating the proposed technique. 
A breakdown of computational time and memory required to solve this problem with the macromodeling approach and FEKO is presented in Tab.~\ref{tab:squarepatch_16by16}. 
All computations were performed with a single thread on a machine equipped with an Intel Xeon E5-2623 v3 processor. 
We observe that the proposed approach is $14$ times faster and requires $8$ times lower memory than FEKO, which uses the MLFMM to simulate the problem.
Since the MLFMM requires accurate computation of near-field interactions, it is not well-suited for multiscale problems with fine mesh size because the cost to compute and store near-field interactions is extremely high. 
In our approach, near-field interactions need to be computed accurately only within a single unit cell, which makes the approach more efficient for multiscale problems.

\begin{table}[t]
\caption{Simulation Statistics for the $16 \times 16$ reflectarray considered in Sec.~\ref{sec:SingleLayerReflectarray}
}
\label{tab:squarepatch_16by16}
\begin{center}
\begin{tabular}{l c  c}
\hline
& FEKO & Proposed\\
\hline
\hline
Total number of unknowns & 479,562 & 177,924\\
Memory used & 307.9~GB & 37.1~GB \\
Macromodel generation & N/A & 10.9~min\\
Matrix fill time & 5.0~h &  14.9~min\\
Preconditioner factorization & 8.7~h & 23.8~min \\
Iterative solver & 1.4~h & 10.8~min\\
Total computation time & 15.2~h & 65~min\\
\hline
\end{tabular}
\end{center}
\end{table}

\subsubsection{${30 \times 30}$ Element Reflectarray}
\label{sec:ticra_30by30}

\begin{figure}[t]
\centering
\subfloat[$\phi = 0^\circ$]{\includegraphics[width=0.95\columnwidth,trim=0.8cm 0.5cm 1cm 1cm,clip=true]{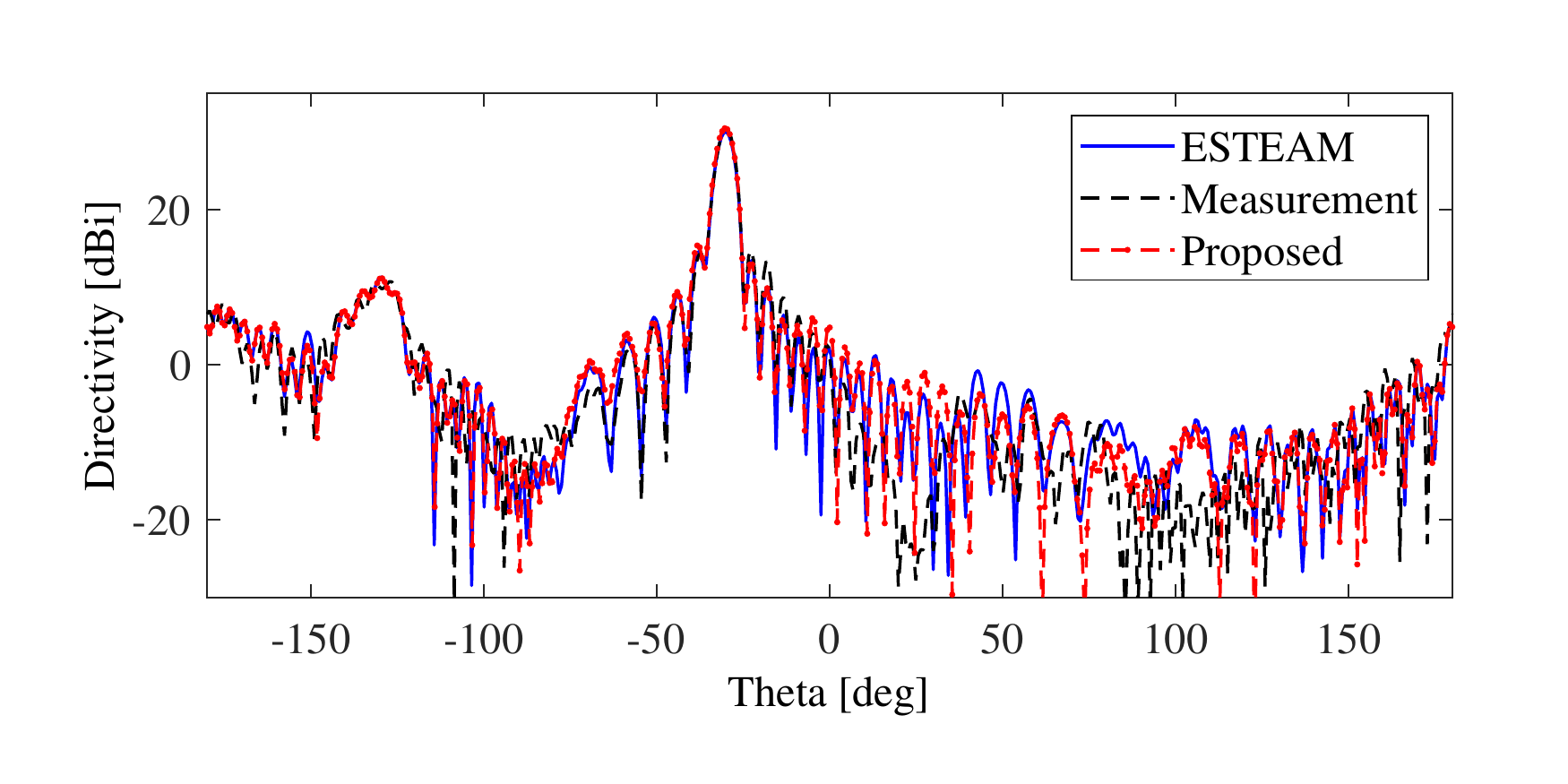}} \\
\subfloat[$\phi = 90^\circ$]{\includegraphics[width=0.95\columnwidth,trim=0.8cm 0.5cm 1cm 1cm,clip=true]{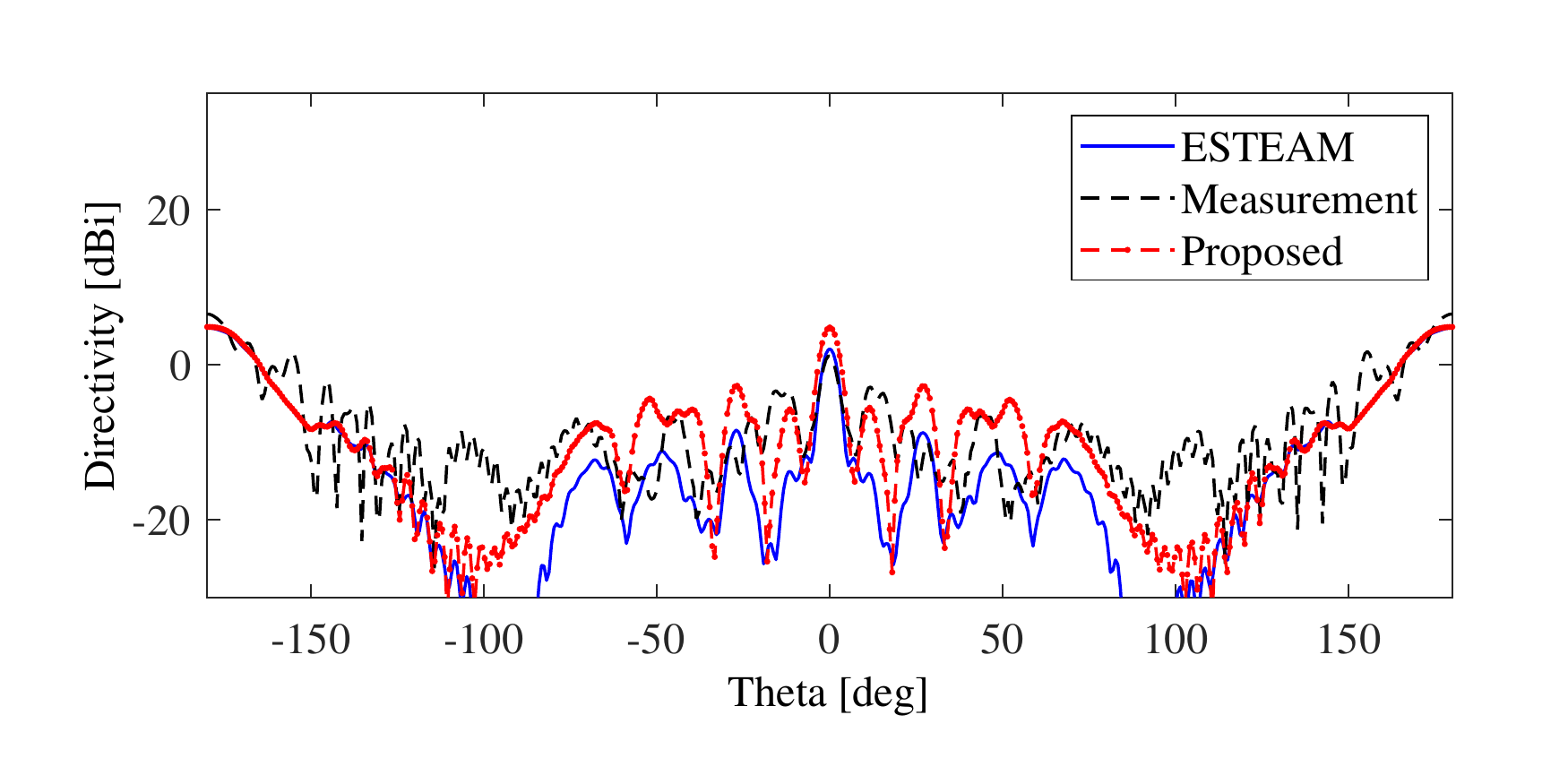}} \\
\subfloat[$\phi = 45^\circ$]{\includegraphics[width=0.95\columnwidth,trim=0.8cm 0.5cm 1cm 1cm,clip=true]{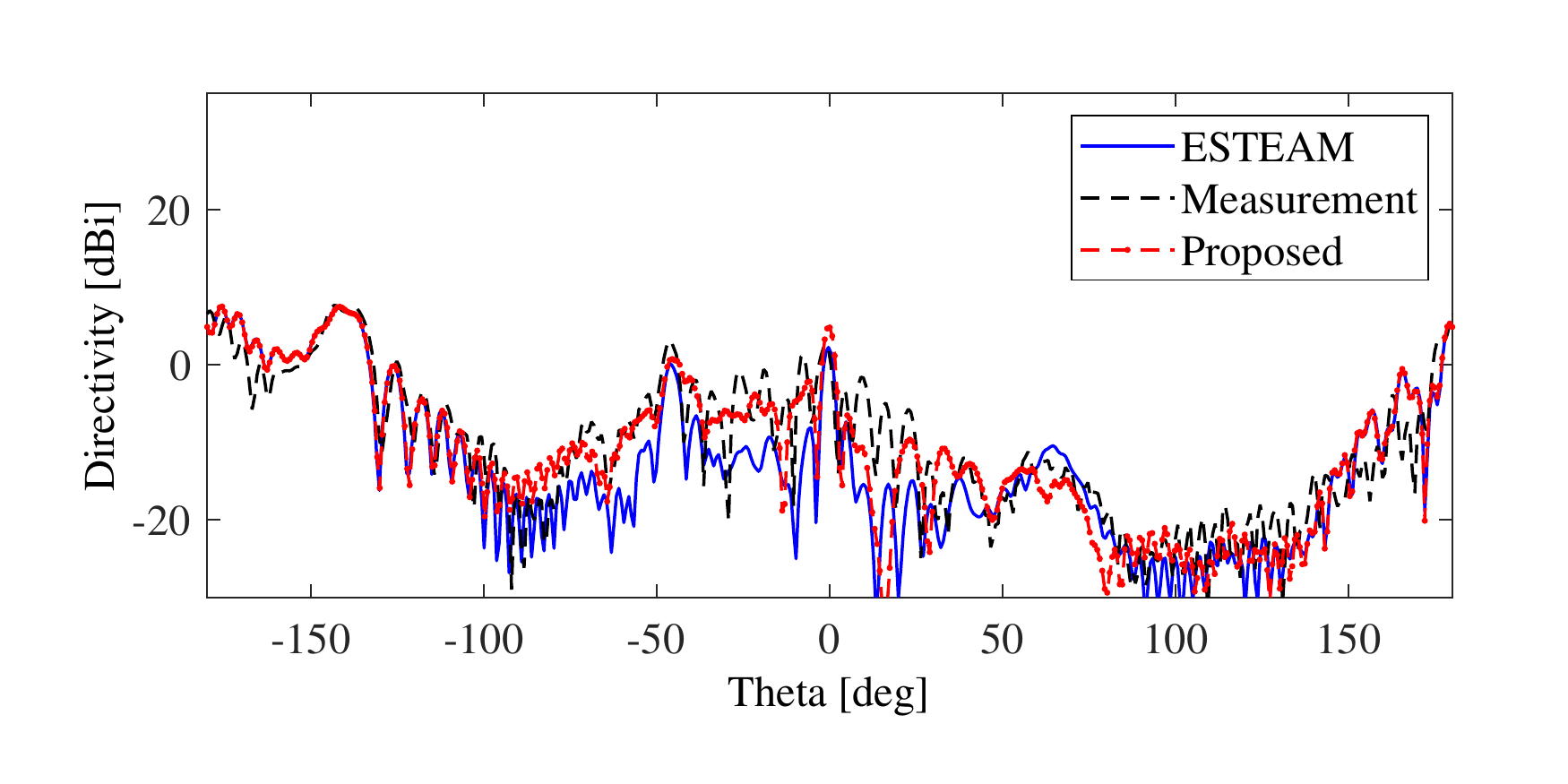}}
\caption{Directivity of the $30 \times 30$ reflectarray considered in Sec.~\ref{sec:ticra_30by30} calculated with TICRA MLFMM solver~\cite{TICRA}, measurement, and the proposed technique.}
\label{fig:directivity_30by30}
\end{figure}

Now, let us consider the reflectarray of size $30 \times 30$. This reflectarray is composed of $32$ distinct unit cells.
In the proposed macromodeling approach, we used the same mesh settings as the $16 \times 16$ case. However, it was not possible to simulate the structure with the same mesh settings in FEKO due to insufficient memory. 
Therefore, we validated the results against the MLFMM solver in TICRA ESTEAM~\cite{TICRA}.
Furthermore, we also compare the results against an experimental results presented previously~\cite{Zhou2015}.
The directivity obtained with the proposed macromodeling approach,  TICRA ESTEAM, and measurements~\cite{Zhou2015} in the $\phi = 0^{\circ}$, $\phi = 90^{\circ}$, and $\phi = 45^{\circ}$ cuts is shown in Fig.~\ref{fig:directivity_30by30}. 
We observe that all three curves match well, further validating the proposed technique.

The simulation with the proposed technique was run on a machine with an Intel Xeon E5-2623 v3 processor, while the simulation with TICRA ESTEAM was run on a machine with   two Intel Xeon  E5-2670 processors (a total of 20 cores).
``High'' accuracy setting was used to simulate this structure in ESTEAM. 
Simulation statistics with the proposed macromodeling approach and TICRA ESTEAM for this test case are summarized in Tab.~\ref{tab:squarepatch_30by30}. 
As summarized in Tab.~\ref{tab:squarepatch_30by30}, the proposed macromodeling approach took $4.4~\mathrm{h}$ to simulate this structure on a single thread. 
TICRA ESTEAM solver took $1.57~\mathrm{h}$ to simulate this structure using 20 cores. 
For a fair comparison, we also present the CPU time in TICRA ESTEAM if the simulation were run using a single thread.
This time is computed assuming a 55\% parallelization efficiency~\cite{Priv_Comm_MinZou}.
For this simulation, the proposed technique which was implemented with double precision arithmetic required 143.8~GB memory, while TICRA ESTEAM solver using single-precision arithmetic required 131~GB memory. 
The ability of the proposed technique to accurately simulate such a large structure highlights the potential of the proposed macromodeling technique.

\subsection{A Two-Layer Reflectarray with Jerusalem Crosses}
\label{sec:twolayer_reflectarray}

\begin{table}[t]
\caption{Simulation Statistics for the $30 \times 30$ reflectarray considered in Sec.~\ref{sec:SingleLayerReflectarray}
}
\label{tab:squarepatch_30by30}
\begin{center}
\begin{tabular}{l c c  c }
\hline
& ESTEAM-I & ESTEAM-II & Proposed  \\
\hline
\hline
Total number of unknowns & - & - & 702,468  \\
Memory used & 131~GB & 131~GB & 143.8~GB \\
Macromodel generation & - & -  & 0.35~h \\
Matrix fill time & - & - & 1.20~h \\  
Preconditioner factorization & - & - & 2.1~h  \\
Iterative solver & - & - & 0.67~h \\
No. of cores & 20 & 1 & 1 \\
Total computation time & 1.57~h & 17.2~h* & 4.4~h\\ 
\hline
\end{tabular}
* estimated based on 55\% parallelization efficiency
\end{center}
\end{table}

\begin{figure}[t]
\centering
\pgfmathsetmacro{\UnitcellWidth}{3.75}
\pgfmathsetmacro{\UnitcellLength}{3.75}
\pgfmathsetmacro{\UnitcellHeight}{3.75}
\pgfmathsetmacro{\ldyl}{3.0}
\pgfmathsetmacro{\ldxl}{3.0}
\pgfmathsetmacro{\tgdcx}{1.6}
\pgfmathsetmacro{\tgdcy}{1.6}
\pgfmathsetmacro{\lcxl}{\ldyl-\tgdcx}
\pgfmathsetmacro{\lcyl}{\ldxl-\tgdcy}
\pgfmathsetmacro{\wdxl}{0.5}
\pgfmathsetmacro{\wdyl}{\wdxl}
\pgfmathsetmacro{\wcxl}{0.3}
\pgfmathsetmacro{\wcyl}{\wcxl}
\pgfmathsetmacro{\h}{0.76}
\pgfmathsetmacro{\hTwo}{-0.76}

\begin{tikzpicture}[xscale= 0.4]\begin{scope}[every node/.append style={xslant=3},xslant=3, xscale= 1, yscale = 0.15]
\coordinate (p1) at (-\UnitcellWidth*0.5,-\UnitcellLength*0.5);
\coordinate (p2) at (\UnitcellWidth*0.5,-\UnitcellLength*0.5);
\coordinate (p3) at (\UnitcellWidth*0.5, \UnitcellLength*0.5);
\coordinate (p4) at (-\UnitcellWidth*0.5,\UnitcellLength*0.5);
\draw[fill=brown!90] (p1)--(p2)--(p3)--(p4)--(p1);
\end{scope}
\node at (-\UnitcellWidth*0.1,-\UnitcellLength*0.15) {\footnotesize $w$};
\node at (\UnitcellWidth*0.7,-\UnitcellLength*0.02) {\footnotesize $w$};
\draw[<->] (-\UnitcellWidth*0.68, -\UnitcellWidth*0.1) --(\UnitcellWidth*0.28, -\UnitcellWidth*0.1);

\draw[<->] (\UnitcellWidth*0.80, \UnitcellWidth*0.078) --(\UnitcellWidth*0.31, -\UnitcellWidth*0.085);

\draw[<->](\UnitcellWidth*0.80, \h*0.4) -- (\UnitcellWidth*0.8, \h*1.3);

\draw[<->](\UnitcellWidth*0.80, \h*1.4) -- (\UnitcellWidth*0.8, \h*2.3);

\node at (\UnitcellWidth*0.8, \h*0.85) [right] {\footnotesize $h$};
\node at (\UnitcellWidth*0.8, \h*1.85) [right] {\footnotesize $h$};

\draw[fill=yellow!40!white, opacity=0.85] (p1) -- (p2) --++(0,\h) --++(-\UnitcellWidth,0) -- (p1);


\begin{scope}[shift = {($(0,\h)$)}, every node/.append style={xslant=3},xslant=3, xscale= 1, yscale = 0.15]

\coordinate (p1) at (-\UnitcellWidth*0.5,-\UnitcellLength*0.5);
\coordinate (p2) at (\UnitcellWidth*0.5,-\UnitcellLength*0.5);
\coordinate (p3) at (\UnitcellWidth*0.5, \UnitcellLength*0.5);
\coordinate (p4) at (-\UnitcellWidth*0.5,\UnitcellLength*0.5);
\draw[fill=yellow!40!white, opacity=0.85] (p1)--(p2)--(p3)--(p4)--(p1);

\coordinate (pp+1) at ($(-0.5*\lcxl, 0.5*\ldyl )$);
\coordinate (pp+2) at ($(\lcxl*0.5, \ldyl*0.5)$);
\coordinate (pp+3) at ($(\lcxl*0.5, \ldyl*0.5 - \wcyl)$);
\coordinate (pp+4) at ($(\wdxl*0.5, \ldyl*0.5 - \wcyl)$);
\coordinate (pp+5) at ($(\wdxl*0.5, \wdyl*0.5)$);
\coordinate (pp+6) at ($(\ldxl*0.5 - \wcxl, \wdyl*0.5)$);
\coordinate (pp+7) at ($(\ldxl*0.5 - \wcxl, \lcyl*0.5)$);
\coordinate (pp+8) at ($(\ldxl*0.5, \lcyl*0.5)$);
\coordinate (pp+9) at ($(\ldxl*0.5, -\lcyl*0.5)$);
\coordinate (pp+10) at ($(\ldxl*0.5- \wcxl, -\lcyl*0.5)$);
\coordinate (pp+11) at ($(\ldxl*0.5 - \wcxl,  -\wdyl*0.5)$);
\coordinate (pp+12) at ($(\wdxl*0.5, -\wdyl*0.5)$);
\coordinate (pp+13) at ($(\wdxl*0.5, -\ldyl*0.5 + \wcyl)$);
\coordinate (pp+14) at ($(\lcxl*0.5,  -\ldyl*0.5 + \wcyl)$);
\coordinate (pp+15) at ($(\lcxl*0.5, -\ldyl*0.5)$);
\coordinate (pp+16) at ($(-\lcxl*0.5, -\ldyl*0.5)$);
\coordinate (pp+17) at ($(-\lcxl*0.5, -\ldyl*0.5+ \wcyl)$);
\coordinate (pp+18) at ($(-\wdxl*0.5, -\ldyl*0.5+\wcyl)$);
\coordinate (pp+19) at ($(-\wdxl*0.5,  -\wdyl*0.5)$);
\coordinate (pp+20) at ($(-\ldxl*0.5 + \wcxl, -\wdyl*0.5)$);
\coordinate (pp+21) at ($(-\ldxl*0.5 + \wcxl,  -\lcyl*0.5)$);
\coordinate (pp+22) at ($(-\ldxl*0.5,  -\lcyl*0.5)$);
\coordinate (pp+23) at ($(-\ldxl*0.5, \lcyl*0.5)$);
\coordinate (pp+24) at ($(-\ldxl*0.5+\wcxl, \lcyl*0.5)$);
\coordinate (pp+25) at ($(-\ldxl*0.5+\wcxl,  \wdyl*0.5)$);
\coordinate (pp+26) at ($(-\wdxl*0.5,  \wdyl*0.5)$);
\coordinate (pp+27) at ($(-\wdxl*0.5,  \ldyl*0.5-\wcyl)$);
\coordinate (pp+28) at ($(-\lcxl*0.5,  \ldyl*0.5-\wcyl)$);

\draw[fill = brown!70] (pp+1) -- (pp+2) -- (pp+3) -- (pp+4) -- (pp+5) -- (pp+6) -- (pp+7) -- (pp+8) -- (pp+9) --(pp+10) --(pp+11) --(pp+12) --(pp+13) --(pp+14) --(pp+15) --(pp+16) --(pp+17) --(pp+18) --(pp+19) --(pp+20) --(pp+21) --(pp+22) --(pp+22) --(pp+23) --(pp+24) --(pp+25) --(pp+26) --(pp+27) --(pp+28) --(pp+1);
\end{scope}
\draw[fill=yellow!40!white, opacity=0.85] (p2) -- (p3) --++(0,-\h) -- ($(p2) + (0,-\h)$) -- (p2);

\draw[fill=green!40!white, opacity=0.55] (p1) -- (p2) --++(0,\h) --++(-\UnitcellWidth,0) -- (p1);
\begin{scope}[shift = {($(0,2*\h)$)}, every node/.append style={xslant=3},xslant=3, xscale= 1, yscale = 0.15]

\coordinate (p1) at (-\UnitcellWidth*0.5,-\UnitcellLength*0.5);
\coordinate (p2) at (\UnitcellWidth*0.5,-\UnitcellLength*0.5);
\coordinate (p3) at (\UnitcellWidth*0.5, \UnitcellLength*0.5);
\coordinate (p4) at (-\UnitcellWidth*0.5,\UnitcellLength*0.5);
\draw[fill=green!40!white, opacity=0.55] (p1)--(p2)--(p3)--(p4)--(p1);

\coordinate (pp+1) at ($(-0.5*\lcxl, 0.5*\ldyl )$);
\coordinate (pp+2) at ($(\lcxl*0.5, \ldyl*0.5)$);
\coordinate (pp+3) at ($(\lcxl*0.5, \ldyl*0.5 - \wcyl)$);
\coordinate (pp+4) at ($(\wdxl*0.5, \ldyl*0.5 - \wcyl)$);
\coordinate (pp+5) at ($(\wdxl*0.5, \wdyl*0.5)$);
\coordinate (pp+6) at ($(\ldxl*0.5 - \wcxl, \wdyl*0.5)$);
\coordinate (pp+7) at ($(\ldxl*0.5 - \wcxl, \lcyl*0.5)$);
\coordinate (pp+8) at ($(\ldxl*0.5, \lcyl*0.5)$);
\coordinate (pp+9) at ($(\ldxl*0.5, -\lcyl*0.5)$);
\coordinate (pp+10) at ($(\ldxl*0.5- \wcxl, -\lcyl*0.5)$);
\coordinate (pp+11) at ($(\ldxl*0.5 - \wcxl,  -\wdyl*0.5)$);
\coordinate (pp+12) at ($(\wdxl*0.5, -\wdyl*0.5)$);
\coordinate (pp+13) at ($(\wdxl*0.5, -\ldyl*0.5 + \wcyl)$);
\coordinate (pp+14) at ($(\lcxl*0.5,  -\ldyl*0.5 + \wcyl)$);
\coordinate (pp+15) at ($(\lcxl*0.5, -\ldyl*0.5)$);
\coordinate (pp+16) at ($(-\lcxl*0.5, -\ldyl*0.5)$);
\coordinate (pp+17) at ($(-\lcxl*0.5, -\ldyl*0.5+ \wcyl)$);
\coordinate (pp+18) at ($(-\wdxl*0.5, -\ldyl*0.5+\wcyl)$);
\coordinate (pp+19) at ($(-\wdxl*0.5,  -\wdyl*0.5)$);
\coordinate (pp+20) at ($(-\ldxl*0.5 + \wcxl, -\wdyl*0.5)$);
\coordinate (pp+21) at ($(-\ldxl*0.5 + \wcxl,  -\lcyl*0.5)$);
\coordinate (pp+22) at ($(-\ldxl*0.5,  -\lcyl*0.5)$);
\coordinate (pp+23) at ($(-\ldxl*0.5, \lcyl*0.5)$);
\coordinate (pp+24) at ($(-\ldxl*0.5+\wcxl, \lcyl*0.5)$);
\coordinate (pp+25) at ($(-\ldxl*0.5+\wcxl,  \wdyl*0.5)$);
\coordinate (pp+26) at ($(-\wdxl*0.5,  \wdyl*0.5)$);
\coordinate (pp+27) at ($(-\wdxl*0.5,  \ldyl*0.5-\wcyl)$);
\coordinate (pp+28) at ($(-\lcxl*0.5,  \ldyl*0.5-\wcyl)$);

\draw[fill = brown!70] (pp+1) -- (pp+2) -- (pp+3) -- (pp+4) -- (pp+5) -- (pp+6) -- (pp+7) -- (pp+8) -- (pp+9) --(pp+10) --(pp+11) --(pp+12) --(pp+13) --(pp+14) --(pp+15) --(pp+16) --(pp+17) --(pp+18) --(pp+19) --(pp+20) --(pp+21) --(pp+22) --(pp+22) --(pp+23) --(pp+24) --(pp+25) --(pp+26) --(pp+27) --(pp+28) --(pp+1);
\end{scope}
\draw[fill=green!40!white, opacity=0.55] (p2) -- (p3) --++(0,-\h) -- ($(p2) + (0,-\h)$) -- (p2);
\end{tikzpicture}
\caption{Unit cell of the two-layer reflectarray considered in Sec.~\ref{sec:twolayer_reflectarray} with $w = 10~\mathrm{mm}$, $h = 0.762~\mathrm{mm}$. Reflectarray has a two-layer dielectric substrate (shown in green and yellow). The top layer has relative permittivity of $\varepsilon_{r} = 3.0$ and the bottom layer has relative permittivity of $\varepsilon_{r} = 2.2$. The reflectarray is backed by a PEC ground plane.}
\label{fig:Unitcell_JC}
\end{figure}
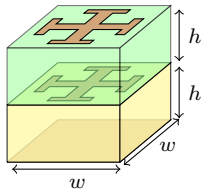

\begin{figure}[t]
\centering
\includegraphics[width = 0.3\textwidth]{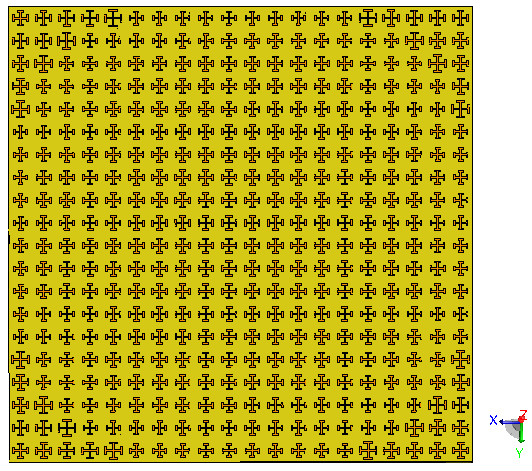} 
\caption{Top view of patch reflectarrays of various sizes considered in Sec.~\ref{sec:twolayer_reflectarray}.}
\label{fig:Topview_JC}
\end{figure}

We now consider a $20 \times 20$ two-layer dual-polarized reflectarray made up of Jerusalem crosses~\cite{Geaney2017}. The top view of this reflectarray is shown in Fig.~\ref{fig:Topview_JC}. 
This example was chosen to demonstrate that the proposed macromodeling approach can simulate electromagnetic surfaces with multiple layers and more complex unit cells as compared to the example in Sec.~\ref{sec:SingleLayerReflectarray}. The reflectarray is composed of $11$ distinct unit cells.
Each unit cell has dimensions of $10~\mathrm{mm} \times 10~\mathrm{mm}$. A sample unit cell is shown in Fig.~\ref{fig:Unitcell_JC}. The reflectarray substrate has two layers, each with a thickness of $0.762~\mathrm{mm}$. The relative permittivity of the bottom and top layers of the substrate is $\varepsilon_r = 2.2$ and $\varepsilon_r = 3.0$, respectively.
The reflectarray is center fed by a horn antenna operating at $f = 10~\mathrm{GHz}$, which is modeled with a spherical wave expansion of a measured horn. 
The horn antenna is placed $0.4~\mathrm{m}$  away from the reflectarray along its axis (f/D = 0.5).

Analysis of this reflectarray example constitutes a multiscale problem. The structure has dimensions of $6.66 \lambda_0 \times 6.66 \lambda_0$, where $\lambda_0$ is wavelength in free space, while each unit cell is only $\lambda_0/3 \times \lambda_0/3$ wide. Furthermore, the size of Jerusalem crosses in each unit cell is between $\lambda_0/5$ to $\lambda_0/4$, while their widths are approximately $\lambda_0/15$. 
Simulation of a reflectarray of this size and complexity is difficult, if not impossible, with existing integral equation solvers.
As such, due to insufficient memory, we could not simulate this reflectarray in FEKO using the MLFMM solver on a $256~\mathrm{GB}$ machine. 
However, the proposed macromodeling solver was able to simulate this $20 \times 20$ reflectarray using $129.3~\rm{GB}$ of memory. 
This was only possible because in the proposed method only $553$ unknowns were required for each unit cell, as opposed to the $6,847$ unknowns (on average) required with the traditional surface integral equation method based on the PMCHWT formulation. 
Overall, this meant that a total of $329,368$ unknowns had to be solved with the proposed macromodeling technique, instead of an estimated $3,080,000$ unknowns by the PMCHWT formulation. 
Simulation of this reflectarray took $10.3~\rm{h}$ on a single thread with the proposed macromodeling approach. 

\begin{figure}[t]
\centering
\subfloat[$\phi = 0^\circ$]{\includegraphics[width=0.95\columnwidth]{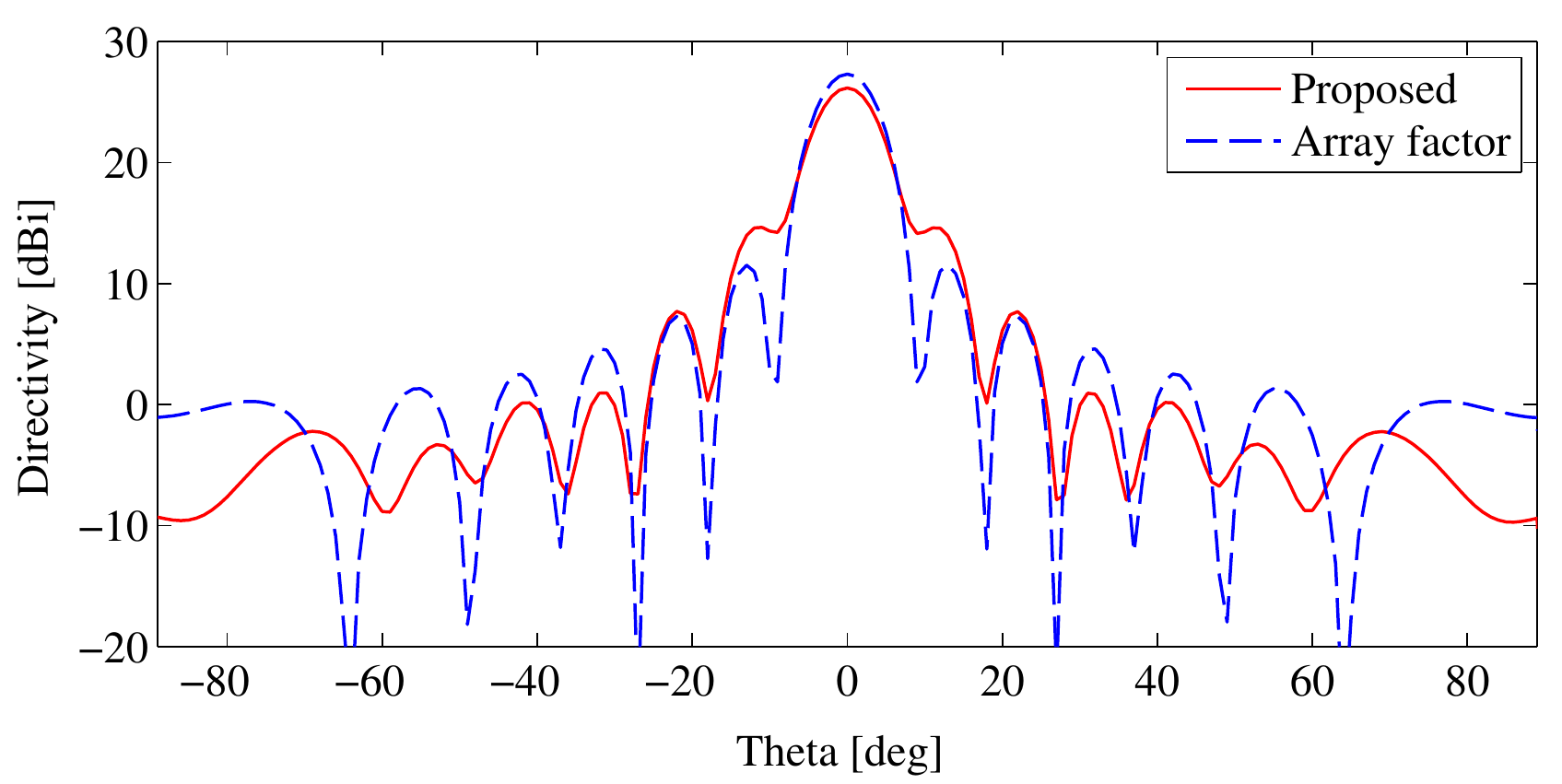}} \\
\subfloat[$\phi = 90^\circ$]{\includegraphics[width=0.95\columnwidth]{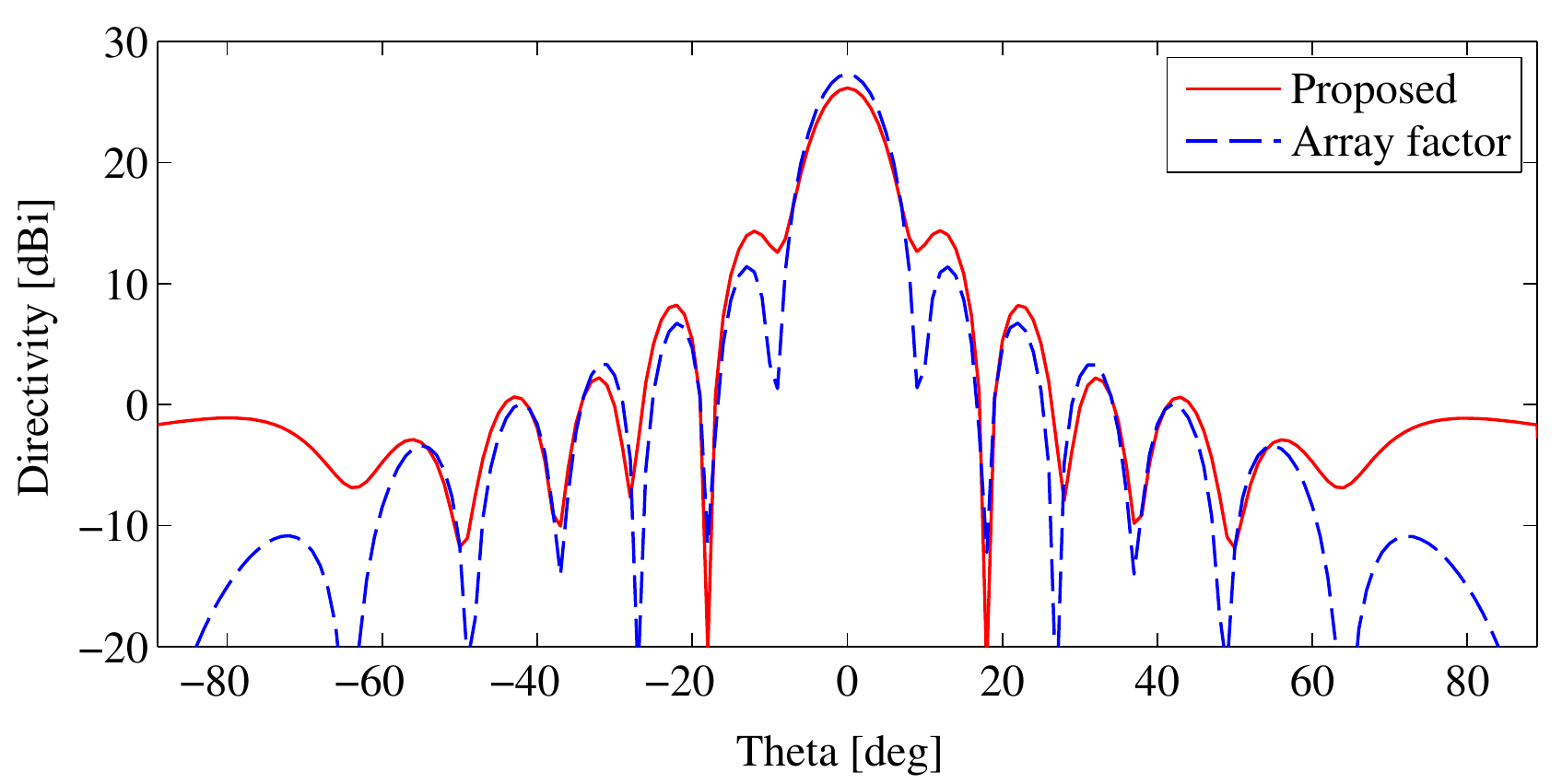}} \\
\caption{Directivity of the $20\times 20$ reflectarray considered in Sec.~\ref{sec:twolayer_reflectarray}.}
\label{fig:Directivity_JC_Reflectarray}
\end{figure}

The reflectarray was designed to radiate the main beam in the broadside direction. 
Figure~\ref{fig:Directivity_JC_Reflectarray} shows the scattered field directivity  of the reflectarray in the $\phi = 0^\circ$, $\phi = 90^\circ$, and $\phi = 45^\circ$ cuts obtain with the proposed technique and periodic analysis based on array factor calculations~\cite{Huang2005}. 
These results demonstrate that full-wave solvers are necessary for an accurate prediction of directivity. 
While periodic analysis can correctly predict the main beam direction, it does not predict properly side lobe levels, null locations, or maximum directivity correctly.

\section{Conclusions}
\label{sec:conclusion}
In this paper, we investigated whether or not it is possible to simulate electromagnetic surfaces, such as reflectarrays and metasurfaces, with an array of macromodels, where  each macromodel captures the scattering behavior of a single unit cell. 
Through the application of the equivalence principle, the Stratton-Chu formulation, and the Schur complement, we demonstrated that indeed the scattering behavior of a complex unit cell can be fully captured via a macromodel operator and equivalent electric and magnetic current densities on a fictitious surface enclosing the scatterer.
In particular, we demonstrated how to tackle problems where fictious surfaces traverse a multilayer dielectric substrate, how to enforce boundary conditions when two or more fictitious surfaces partially overlap, and how to model ground planes that coincide with fictitious surfaces.
The proposed macromodeling approach helps restore the periodicity of the problem by turning an array of heterogeneous scatterers into an array of equivalent current densities on a periodic mesh. This property allows us to rigorously capture the mutual coupling between array elements via the FFT.
Through numerical examples, we demonstrated that the proposed approach, in terms of accuracy, compares well against both experimental results and simulation results from other full-wave EM solvers.
In terms of computational efficiency, the proposed technique can be up to 14 times faster and can require up to 10 times less memory than commercial MLFMM solvers.
Ultimately, the proposed macromodeling approach could allow simulations of complex EM surfaces that are not feasible with existing approaches due to excessive memory consumption or computation times.

\section{Acknowledgment}

Authors would like to thank Dr. Min Zhou from TICRA for providing simulation and experimental results for the test case in Sec.~\ref{sec:SingleLayerReflectarray}.

\bibliographystyle{IEEEtran}
\bibliography{IEEEabrv,biblio}

\end{document}